\documentclass{aa}
\usepackage{graphicx}
\usepackage{txfonts}
\usepackage{general}
\usepackage[version=3]{mhchem}
\begin{document}
\newcommand\nratio{{$^{14}$N/$^{15}$N}}
\newcommand{\rnow}{\ensuremath{{\cal R}_0}}
\newcommand{\rsun}{\ensuremath{{\cal R}_\odot}}
\newcommand{\rr}{\ensuremath{{\cal R}}}
\newcommand{\rrs}[1]{\ensuremath{{\cal R}_{\rm #1}}}

\title{Direct evidence of multiple reservoirs of volatile nitrogen in
  a protosolar nebula analogue.}  \titlerunning{Multiple reservoirs of
  nitrogen in a protosolar nebula analogue}

\author{%
  P. Hily-Blant\inst{1,2}\and V. Magalhaes\inst{2}\and
  J. Kastner\inst{2,3}\and A. Faure\inst{2}\and T.
  Forveille\inst{2}\and C. Qi\inst{4} } \institute{%
  Institut Universitaire de France
  \email{pierre.hily-blant@univ-grenoble-alpes.fr}\and Universit\'e
  Grenoble Alpes, CNRS, IPAG, F-38000 Grenoble, France \and Chester
  F. Carlson Center for Imaging Science, School of Physics \&
  Astronomy, and Laboratory for Multiwavelength Astrophysics,
  Rochester Institute of Technology, 54 Lomb Memorial Drive, Rochester
  NY 14623 USA\and Harvard-Smithsonian Center for Astrophysics, 60
  Garden Street, Cambridge, MA 02138, USA}

\date{}

\abstract{%
  Isotopic ratios are keys to understanding the origin and early
  evolution of the solar system in the context of Galactic
  nucleosynthesis. The large range of measured $^{14}$N/$^{15}$N
  isotopic ratios in the solar system reflects distinct reservoirs of
  nitrogen whose origins remain to be determined. We have directly
  measured a C$^{14}$N/C$^{15}$N abundance ratio of 323$\pm$30 in the
  disk orbiting the nearby young star TW Hya. This value, which is in
  good agreement with nitrogen isotopic ratios measured for prestellar
  cores, likely reflects the primary present-day reservoir of nitrogen
  in the solar neighbourhood. These results support models invoking
  novae as primary $^{15}$N sources as well as outward migration of
  the Sun over its lifetime, and suggest that comets sampled a
  secondary, $^{15}$N-rich reservoir during solar system formation.}


\maketitle

\section{Introduction}

Understanding the formation of the solar system is a prerequisite to a
comprehensive theory of our origins, while providing essential clues
for the birth of planetary systems in general. Early solar system
bodies, such as comets and asteroids, provide a detailed view of the
composition of the protosolar nebula (PSN) 4.6 billion years ago
\citep{mumma2011,bockelee2015}. The physical and chemical conditions
prevailing in the PSN were set up in the collapsing proto-Sun, which
were themselves products of the conversion, over a few million years,
of diffuse and mostly atomic interstellar gas into a dense,
gravitationally unstable, molecular prestellar core
\citep{ceccarelli2014ppvi}. A key question from both astrophysical and
planetary science perspectives is to know to what extent the
reservoirs of volatiles (namely, gas and ice) in planetary systems are
of interstellar nature or if chemistry was reset in the PSN at the
epoch of planet formation.

Isotopic ratios are a powerful tool to evince the chemical heritage
preserved during this secular evolution. For example, the large D/H
ratio in terrestrial water {may} indicate that it likely encoded the
interstellar history of the solar system \citep{cleeves2014}. In
contrast, the origin of nitrogen, the most reactive and abundant heavy
nuclei, together with carbon and oxygen, has not been elucidated for
two main reasons: i) the elemental \nratio\ isotopic ratio of nitrogen
in the present-day solar neighbourhood (which we denote \rnow) is
poorly constrained, and ii) the primary repository of nitrogen in star
and planet forming regions is unknown, such that chemical model
predictions are highly uncertain. Constraints on the origin of
nitrogen in the PSN derive from laboratory and in situ analysis of
primordial solar system materials \citep{rubin2015a}, and from
observations of solar-type star-forming regions in the solar
neighbourhood within a few 100~pc \citep{hilyblant2013a}. The primary
reservoir of nitrogen in the PSN had a \nratio\ ratio \rsun=441 as
measured in Jupiter's atmosphere and in the solar wind
\citep{furi2015} but values as low as 50 are observed today in some
primitive cosmomaterials \citep{bonal2010}. Comets, observed in CN,
HCN, and NH$_2$, show a strikingly uniform ratio of $\approx140$
regardless of their orbital parameters \citep[][ and
Fig.~\ref{fig:comets}]{mumma2011,shinnaka2016b}, suggesting that they
recorded a secondary nitrogen reservoir {or that the main reservoir of
  nitrogen has not yet been seen in comets.}

A central question is to establish whether the different reservoirs of
nitrogen that are observed today in the solar system were inherited
from the interstellar phase, or if they result from fractionation
processes within the PSN or in parent bodies
\citep{hilyblant2013a,furi2015}. A first step in addressing this issue
would be to demonstrate the existence of multiple reservoirs in
protoplanetary disks, which serve as analogues for the PSN. This is a
primary objective of the present work.

The origin of nitrogen in the solar system is actually a problem in
both space and time, as it involves the comparison of the composition
of the PSN, formed at some location in the Galaxy 4.6 billion years
ago, to that of present-day star-forming regions at a galactocentric
radius $r_G\approx8.5$~kpc. One key point is the chemical homogeneity
of the local interstellar medium (LISM) within $\sim1$~kpc
\citep{sofia2001}, which makes nearby protoplanetary disks
representative of planetary formation at
$r_G\approx8.5$~kpc. Qualitatively, the elemental \nratio\ in the
Galaxy is expected to decrease with time, and to increase with
increasing galactocentric radius, as a result of stellar
nucleosynthesis \citep{audouze1975,matteucci2012}. Nevertheless,
predictions of Galactic chemical evolution (GCE) models can vary
significantly depending on the specific assumptions concerning the
nuclear processes leading to $^{15}$N, the Galactic star formation
history, and the initial mass function of newly born stars
\citep{romano2003, minchev2013}. Moreover, the birthplace of the Sun
remains poorly known \citep{barbosa2015}. Measuring \rnow\ would
provide the reference value {needed to} establish the heretofore
missing link between the PSN and present-day planet-forming disks
orbiting young stars. The value of \rnow\ is the other aim of the
present study.

However, measuring \rnow\ is challenging because the primary reservoir
of nitrogen in cores, protostars, and protoplanetary
disks---presumably N or N$_2$ (or perhaps icy ammonia in the densest
parts)---cannot be observed directly. Therefore, \rnow\ must be
inferred from trace species (e.g. HCN, NH$_3$, etc) usually with the
aid of chemical models \citep{legal2014,roueff2015}. Moreover,
radiation from $^{15}$N isotopologues is intrinsically weaker than
from the main isotopologue by typically two orders of magnitude
{thus requiring excellent sensitivity}.

It is well established that evolved protoplanetary disks are bright CN
line sources \citep{guilloteau2013} and that, even for the strongest
CN sources among disks, some hyperfine transitions of CN are optically
thin \citep{kastner2015, punzi2015}. This species hence potentially
affords a direct means to infer N isotope ratios in disks, without the
need to resort to assumptions concerning secondary elemental isotopic
ratios or line optical depths.

\section{Observations}

With this as motivation, we have undertaken ALMA observations of the
C$^{14}$N(3-2) and C$^{15}$N(3-2) rotational emission {at 340 and
  330~GHz respectively (see Table~\ref{tab:spectro})} from the
molecule-rich disk orbiting TW~Hya, a nearby (d=59.5(9)
pc\footnote{Here and elsewhere, 1$\sigma$ uncertainties are given
  within brackets in units of the last digit}), nearly pole-on T Tauri
star and disk system that displays exceptionally bright, narrow CN
line emission \citep{kastner2015, teague2016}.

Observations of TW~Hya were performed in December 2014 and April 2015,
using the Atacama Large Millimeter/submillimeter Array (ALMA)
interferometer (proposal 2013.1.00196.S). The Fourier plane was
sampled by 47 antennas, covering baselines from 15 to 343~m. The CN
and C$^{15}$N rotational lines $N=3-2$ were observed simultaneously
with a channel spacing of 61~kHz (or 0.107 km/s effective velocity
resolution at 340~GHz). The rest frequency for the C$^{15}$N(3-2)
hyperfine transitions were taken from the CDMS spectroscopic catalogue
\citep{muller2005}, with 1$\sigma$ uncertainties of 90~kHz. The CDMS
frequencies for the CN(3-2) set of hf lines have much reduced
uncertainties of a few 10~kHz only. Flux calibration was performed
using Callisto as an absolute reference, while bandpass and phase
calibrations were obtained by observing the J1256-0547 and J1037-2934
QSOs respectively. After a first round of calibration using the CASA
software (version 4.2.2), the phase calibration was improved using a
self-calibration procedure based on the continuum, line-free,
emission, from large bandwidth observations performed in parallel to
the spectral line observations. The self-calibrated output was then
applied to both CN and C$^{15}$N. Cross-calibration uncertainties were
mitigated by the simultaneous observation of the two isotopologues,
while the small (3\%) difference in lines frequency ensures nearly
identical Fourier plane coverages, continuum emission, and absolute
intensity calibration. The calibrated data were then exported to the
\textsc{Gildas}\footnote{\url{http://www.iram.fr/IRAMFR/GILDAS}} package
format for imaging and analysis.

\begin{figure}[t]
  \centering
  \includegraphics[width=\hsize]{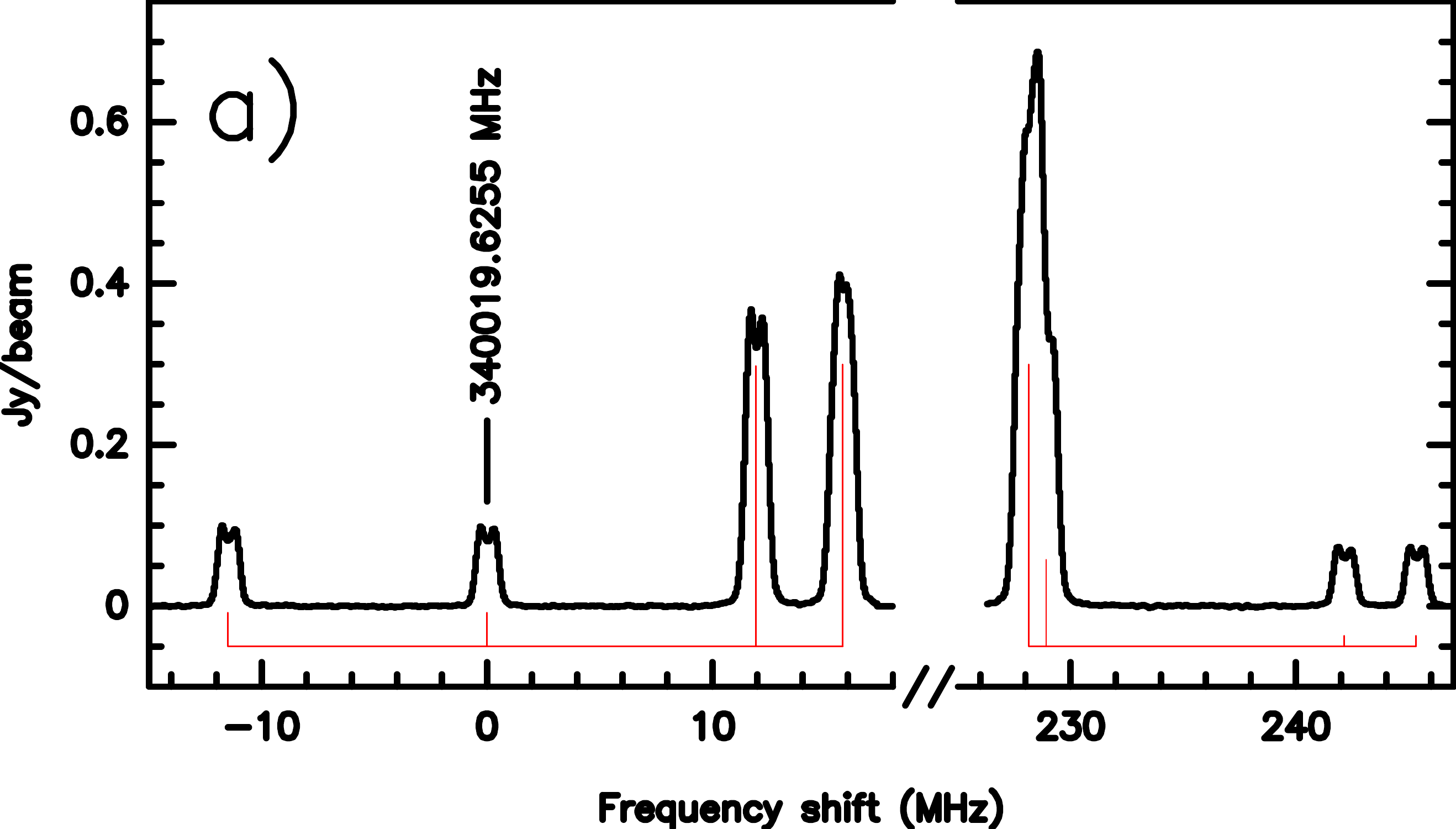}\bigskip\\
  \includegraphics[width=\hsize]{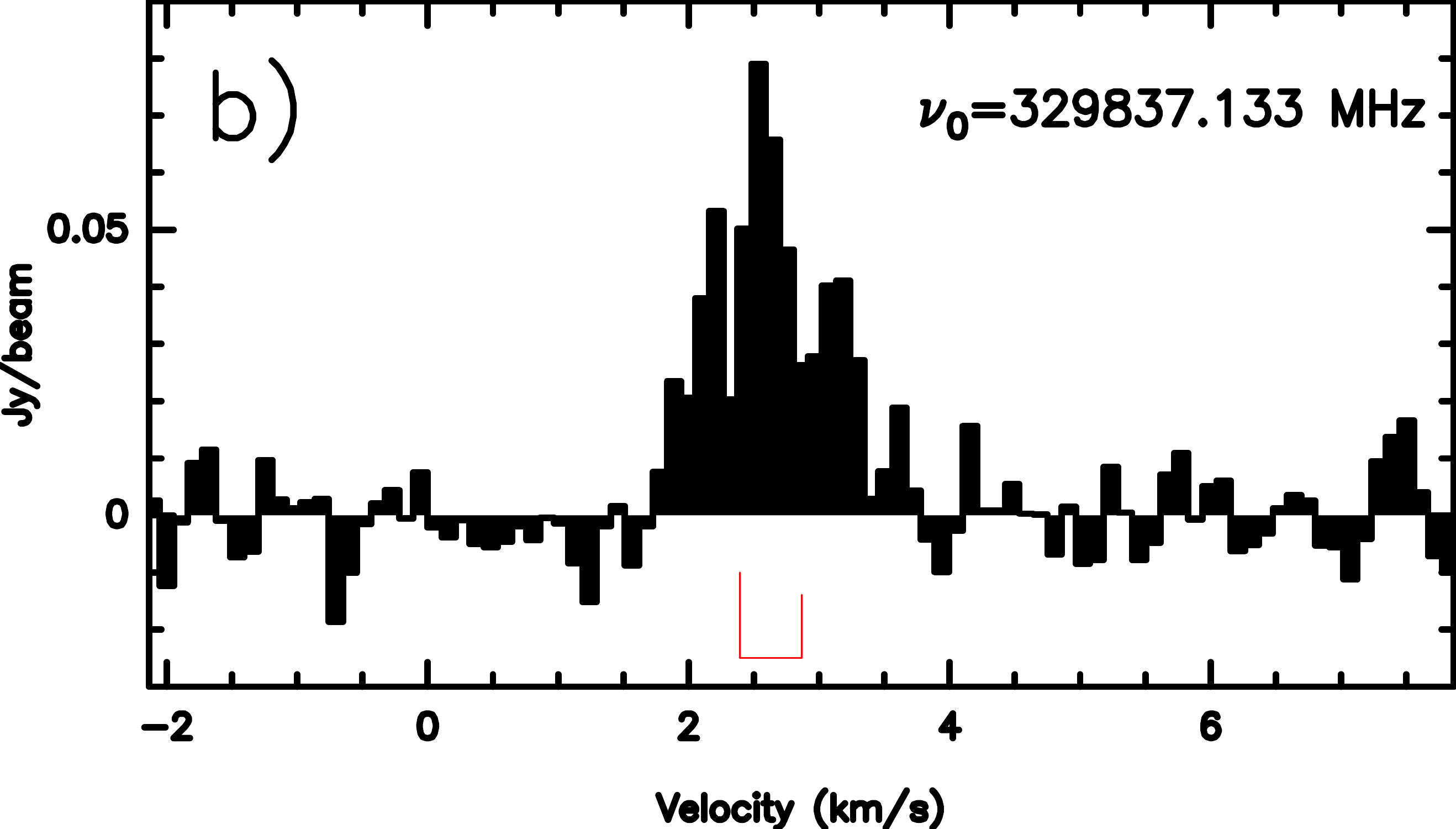}\bigskip\\
  \includegraphics[width=\hsize]{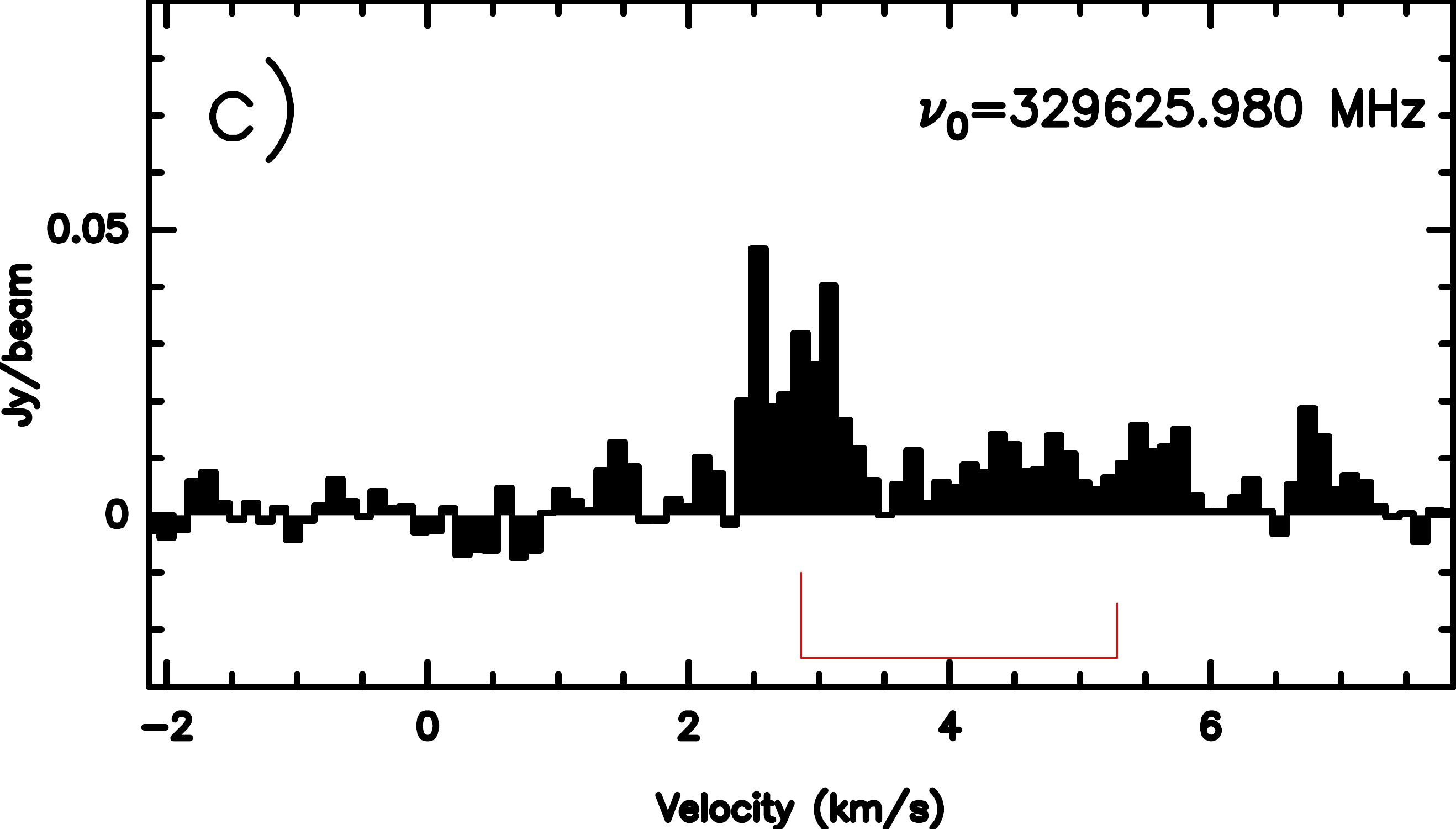}
  \caption{Spectra, averaged within the dashed circle in
    Fig.~\ref{fig:maps}, of the entire C$^{14}$N {(a)} and of
    the two sets of hyperfine lines of C$^{15}$N {(b) and
      (c)}. The hyperfine splitting and relative strengths of the hf
    components are indicated below the spectra.}
\label{fig:spectra}
\end{figure}

\begin{figure}[t]
  \centering
  \includegraphics[width=.9\hsize]{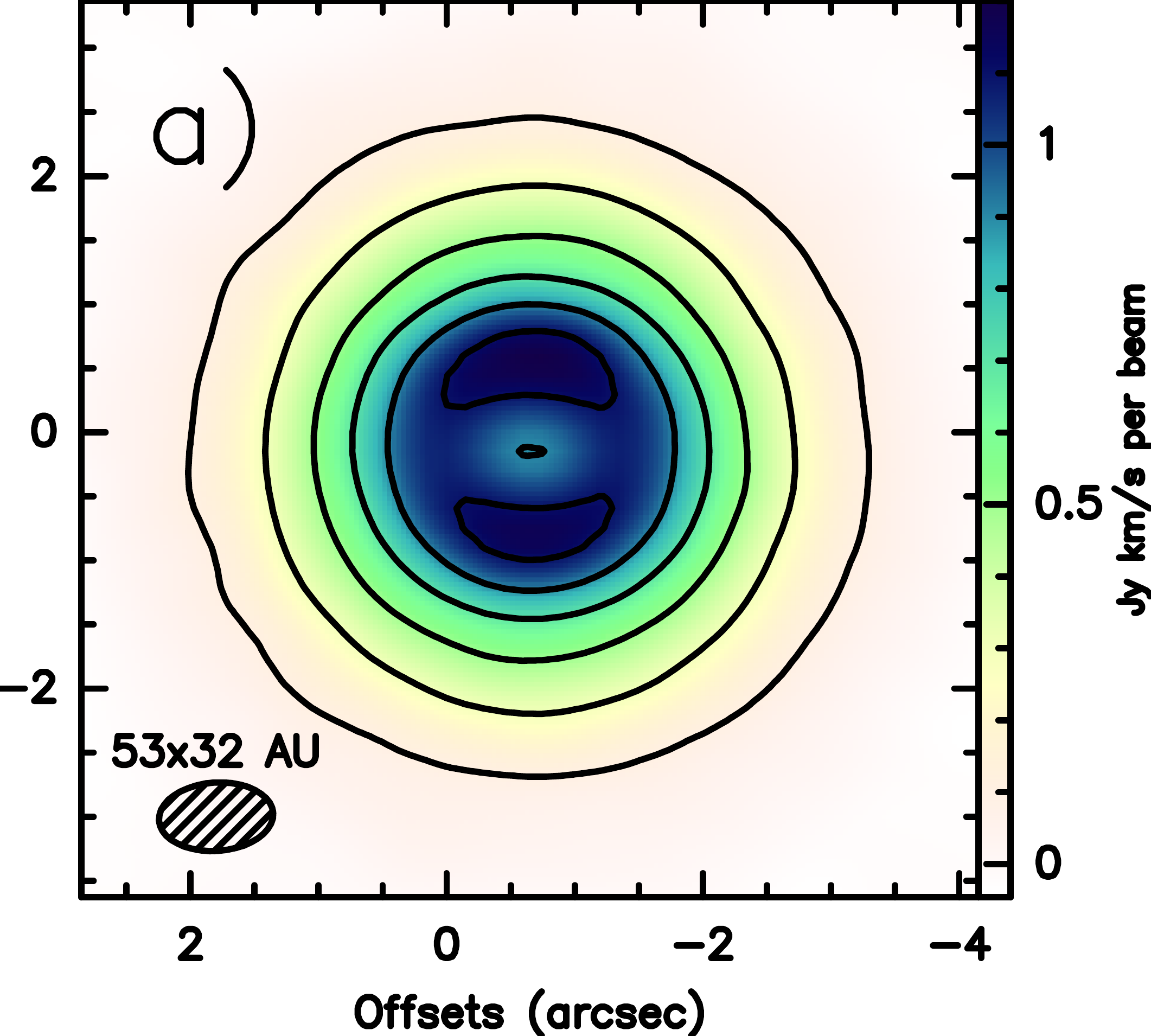}\\
  \includegraphics[width=.9\hsize]{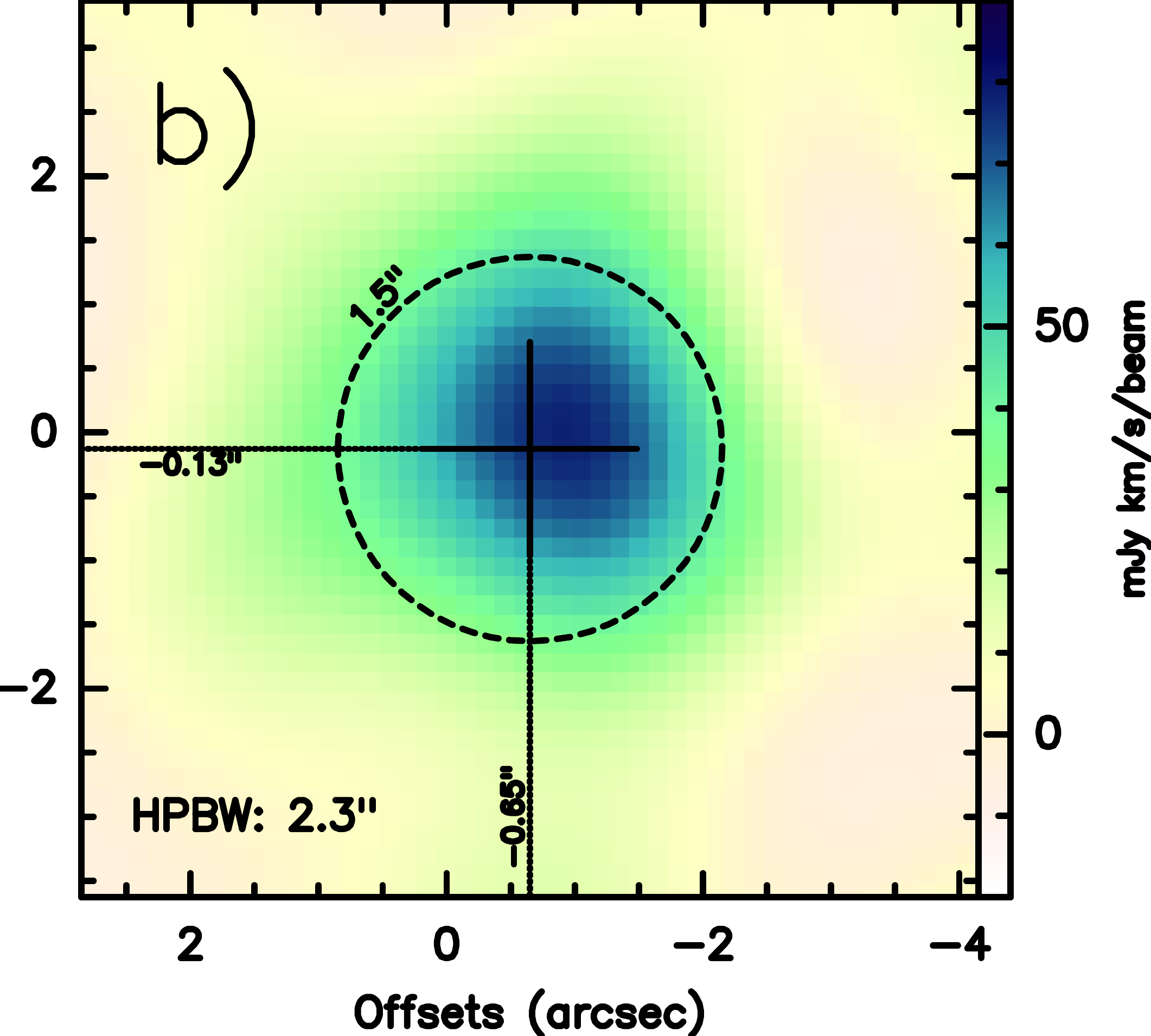}
  \caption{Integrated intensity maps towards TW~Hya with ALMA. a)
    C$^{14}$N emission summed over the four weakest hyperfine
    components. b) C$^{15}$N from 329835.5 to 329838.5~MHz, with
    spatial resolution degraded to 2.3''. The phase center is at
    $11^h01^m51.875^{s}, -34^\circ42'17.16''$ (J2000).}
\label{fig:maps}
\end{figure}

{The CN hyperfine (hf) transitions} are well resolved and detected
with a high signal-to-noise ratio (S/N) (see Figs.~\ref{fig:spectra}
and \ref{fig:maps}).  Specifically, our CN detections include four hf
lines with relative intensities 0.054 and 0.027 (see
Table~\ref{tab:spectro}). Of the four hf components of the
C$^{15}$N($N=$3-2) rotational transition included in our frequency
setups, three were successfully detected and are optically thin, while
the weakest was only tentatively detected {(see panels b and c of
  Fig.~\ref{fig:spectra})}.

To produce the images shown in Fig.~\ref{fig:maps}, deconvolution was
performed using the H\"ogbom algorithm as implemented in the
\textsc{Gildas/Mapping} software. The clean beam at 340~GHz was
0.90''$\times$0.53'' and with a position angle of 86\degr, and the
final sensitivity is between 1.8 and 3.6~mJy/beam/channel. The
continuum peak and integrated flux at 340.0465~GHz are 378.0(3)
mJy/beam and 1.426(2) Jy respectively, in agreement with observations
at nearby frequencies \citep{qi2013,nomura2016}. To increase the S/N
of the C$^{15}$N, robust weighting was applied providing good
compromise between beam size and secondary lobes level. The
synthesized beamwidth is 2.3\arcsec, and the sensitivity was between
5.3 to 6.4 mJy/beam/channel.

\begin{figure}
  \centering
  \includegraphics[width=0.9\hsize]{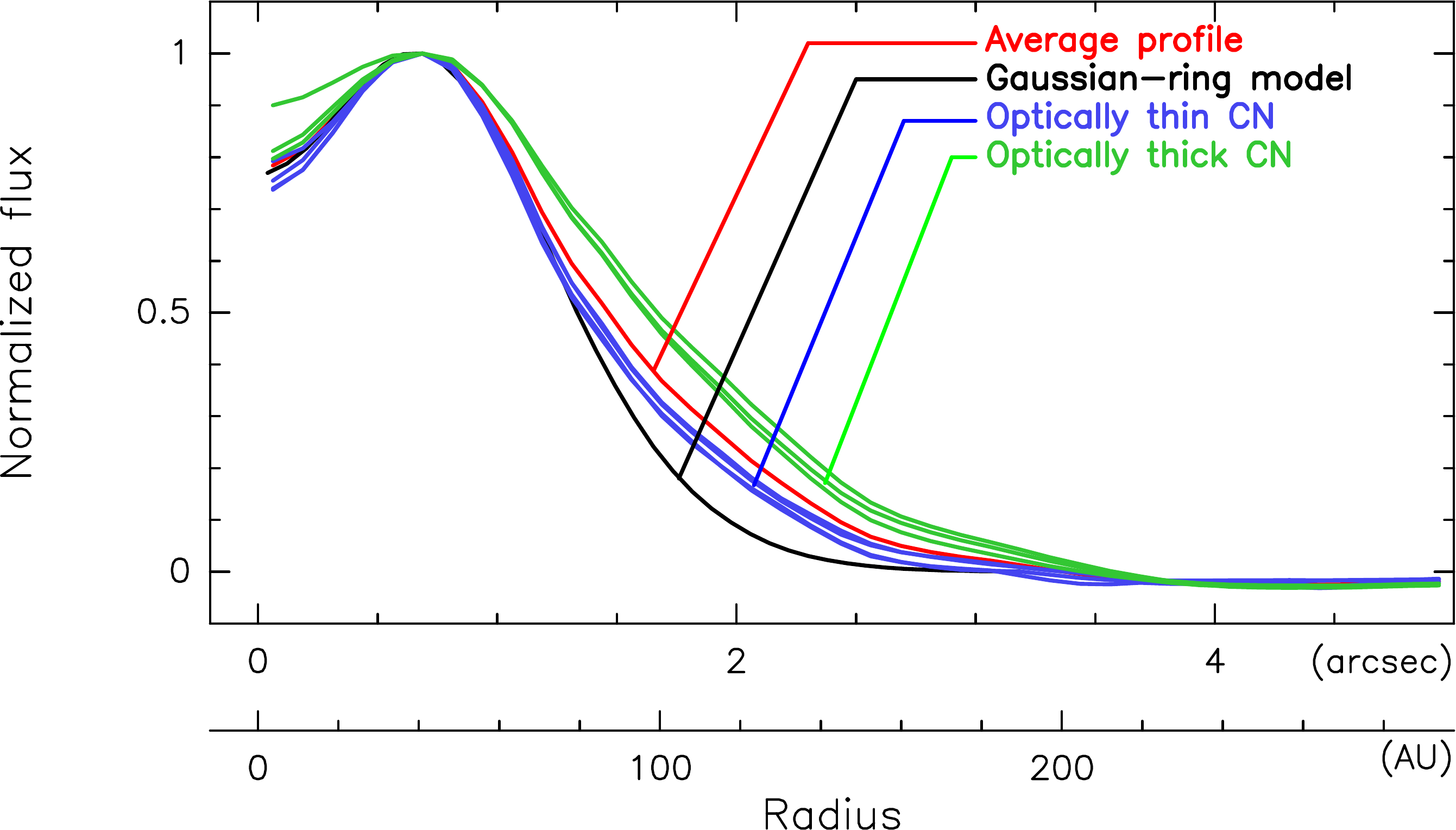}
  \caption{Azimuthally averaged radial profiles of the integrated flux
    for the 7 hyperfine (hf) components of C$^{14}$N, four of which
    are optically thin, the remaining three being optically thick
    (Fig.~\ref{fig:maps}). The average of the 7 hf components is shown
    in red. The black line shows the result of a sharp inner-ring
    model plus Gaussian decrease outwards, convolved by the
    synthesized beam. The {unconvolved} ring has a radius of
    0.7\arcsec\ or 42~AU, and a full-width at half-maximum of 54~AU
    (at a distance of 59.5$\pm$0.9~pc).}
  \label{fig:azimuth}
\end{figure}

The CN map shows a ring structure of deconvolved inner and outer
half-maximum radii of {roughly 15 and 70~AU respectively, as derived
  from a very simple model} (Figs.~\ref{fig:maps} and
\ref{fig:azimuth}), consistent with previous studies based on
molecular and continuum emissions \citep{qi2013, teague2016,
  nomura2016}. The very high S/N channel maps of the optically thin CN
hyperfine transitions at 340.020 and 340.265~GHz are shown in
Fig.~\ref{fig:cmaps}, illustrating both the well-known kinematic
pattern of Keplerian disks and the ring-like emission pattern.

\section{The CN/C\fifn\ abundance ratio in TW Hya}

\begin{table*}[t]
  \begin{center}
    \caption{\label{tab:uvfit} Results of the $uv$-plane analysis (see
      the text for details) using visibilities within a $uv$ radius of
      60~m. The center of the 2D-Gaussian is given with respect to the
      phase center position at RA,DEC coordinates
      $11^h01^m51.875^{s}, 34^\circ42'17.16''$ (J2000). Number in
      brackets are uncertainties in units of the last digit.}
  \begin{tabular}{lcccccr}
    \toprule
    &  \multicolumn{2}{c}{Center position}
    &  \multicolumn{2}{c}{FWHM ('')}
    & Area
    & Flux$^\S$\\
%    & Peak flux\\
    & $\delta$RA ('')
    & $\delta$DEC ('')
    & Major & Minor & arcsec$^2$
    & Jy km/s\\
    \midrule
    % CN
    % & -0.64$\pm$0.01 & -0.13$\pm$0.01 & 2.58$\pm$0.02 & 2.57$\pm$0.02 & 7.50$\pm$0.07 & 6.95$\pm$0.05 & 0.93$\pm$0.01 \\
    % C$^{15}$N  & --- & --- & --- & --- & --- & 9.06$\pm$0.04 & 1.21$\pm$0.01\\
    % With new tables 06-nov-2016
    CN
    & -0.65(1) & -0.13(1) & 2.55(2) & 2.54(2) & 7.33(8) & 2.02(2) \\ % 0.940(10) Jy/(km/s)*2.15
    C$^{15}$N$^\dag$
    & -0.65(0) & -0.13(0) & 2.55(0) & 2.54(0) & 7.33(0) & 0.166(13)\\%0.100(8)*1.66 \\
    C$^{15}$N$^\ddag$
    & -0.71(11) & -0.05(12) & 2.46(35) & 2.13(33) & 5.93(126) & 0.150(20) \\ %0.090(12)*1.66
    % as of 03nov
    % CN
    % & -0.64(1) & -0.13(1) & 2.58(2) & 2.57(2) & 7.50(7) & 6.94(9)\\% & 0.93(1) \\
    % C$^{15}$N$^\dag$
    % & -0.64(0) & -0.13(0) & 2.57(0) & 2.57(0) & 7.50(0) & 8.95(74) \\%& 1.21(1)\\
    % C$^{15}$N$^\ddag$
    % & -0.71(11) & -0.05(12) & 2.46(35) & 2.13(33) & 5.93(126) & 8.02(110)\\
    \bottomrule
  \end{tabular}
  \end{center}
  {\footnotesize {Notes:} $^\S$ Fluxes integrated over the CN(3 7/2
    5/2$\rightarrow$2 5/2 5/2) and the two C$^{15}$N(3 7/2
    $\rightarrow$2 5/2) transitions. The quoted uncertainty does not
    include the calibration uncertainty (3\%).  $^\dag$ $uv$-fit
    results with fixed, circular, Gaussian disk taken from the CN fit.
    $^\ddag$ $uv$-fit results without imposing the size of the
    Gaussian disk.}
  % \begin{list}{}{}
  %   \footnotesize
  % \item[$^\S$] Fluxes integrated over the CN(3 7/2 5/2$\rightarrow$2
  %   5/2 5/2) and the two C$^{15}$N(3 7/2 $\rightarrow$2 5/2)
  %   transitions. The quoted uncertainty does not include the
  %   calibration uncertainty (3\%).
  % \item[$^\dag$] $uv$-fit results with fixed, circular, Gaussian disk
  %   taken from the CN fit.
  % \item[$^\ddag$] indicates the results of a $uv$-fit without imposing
  %   the size of the Gaussian disk.
  % \end{list}
\end{table*}

The disk-averaged CN/C$^{15}$N abundance ratio was derived from the
integrated visibilities of optically thin hf components of each
isotopologue. {The chosen CN hf transition at 340.020~GHz is readily
  shown to be optically thin by considering the total opacity of the
  CN line, which is of order 4--5 \citep{kastner2014}, and recognizing
  that this hf line carries 5.4\% of the total intensity. The low
  optical depth is confirmed by hyperfine analysis of the ALMA spectra
  and by the identical lineshapes of the disk-averaged spectra of the
  4 weakest hf transitions with relative intensities 2.7 and 5.4\%
  (see Fig.~\ref{fig:stacking} and Section~\ref{app:excitation} for
  details). Because C\fifn\ is typically a hundred times (or more)
  less abundant than CN, its detected hf lines are also optically
  thin. The visibilities of the CN hf line and of the overlapping
  C\fifn\ hf transitions at 329.837~GHz were integrated over a 2\kms\
  velocity interval, to optimize S/N, and tapered before fitting in
  the $uv$ plane by a two-dimensional Gaussian distribution. In doing
  so, the geometrical parameters were indeed obtained from the high
  S/N CN map, and then applied to the C$^{15}$N visibilities letting
  the C\fifn\ integrated flux as the unique free parameter.}

{The derived integrated flux ratio was translated into a column
  density ratio by assuming a single excitation temperature for both
  isotopologues. This assumption is strongly supported by the common
  excitation temperature of $\approx20-25$~K for the CN(3-2) (see
  Sec.~\ref{app:excitation}) and CN(2-1) transitions
  \citep{teague2016}. Thermalization is most likely ensured by the
  high density regions responsible for the CN emission, as inferred
  from the low kinetic temperatures traced by the narrow CN lines (see
  Sec.~\ref{app:excitation}).}

After propagating the various sources of uncertainty, the
disk-averaged CN/C$^{15}$N abundance ratio is
\begin{equation}\rrs{CN} = 323(30)\label{eq:cnratio}\end{equation}

\section{Two isotopic reservoirs of nitrogen in disks}

This direct determination of the CN isotopic ratio in TW~Hya can be
compared to the indirectly measured ratio in HCN in the nearby
($d$=140 pc) protoplanetary disk orbiting MWC~480. We have applied the
same $uv$ plane analysis described above to archival H$^{13}$CN and
HC$^{15}$N data for MWC~480 \citep{guzman2015}, obtaining
\begin{equation}{\rm H^{13}CN / HC^{15}N} =
  1.88(34)\label{eq:h13cnratio}\end{equation} which improves on, but
is consistent with, the 2.8(1.4) value formerly derived. The
improvement over the previous analysis is most likely due to our
fitting approach which involves fewer free parameters by assuming
identical disk properties for both isotopologues. Indeed, our ratio
for the MWC~480 disk also agrees well with recent re-analysis of the
data \citep{guzman2017}.

Assuming that the abundance ratio HCN/H$^{13}$CN is equal to the
elemental $^{12}$C/$^{13}$C isotopic ratio, and adopting
{68(15)} for the latter \citep{milam2005}, we obtain an
isotopic ratio of nitrogen in HCN in MWC~480 of
\begin{equation}\rrs{HCN} = 128(36)\label{eq:hcnratio}.\end{equation}

The chemical composition of the solar neighbourhood being homogeneous
\citep{sofia2001}, {it follows that the elemental, hence primary,
  reservoir of nitrogen in both TW~Hya and MWC~480 should have the
  same isotopic ratio}. Consequently, the distinct ratios in CN and
HCN demonstrate that, in addition to the elemental reservoir, at least
one of these two disks contains another reservoir of nitrogen, which
is fractionated and $^{15}$N-rich. {Although the HCN/H\thcn\ in disks
  is not known, the foregoing conclusion would hold true unless this
  ratio was larger than 132. In the context of prestellar cores, such
  high ratios have been predicted by recent chemical models
  \citep{roueff2015}, but are inconsistent with observations which
  lead to a ratio of 30 \citep{daniel2013}.}  Combined with GCE models
predicting that the enrichment in $^{15}$N through stellar
nucleosynthesis over the last 4.6 billion years implies $\rnow>232$
\citep{romano2003}, our result demonstrates that HCN in MWC~480, and
in the disks studied in \cite{guzman2017}, traces a heavily
fractionated, hence secondary, reservoir of nitrogen.

From our study, it appears that CN and HCN in disks are not tracing
the same reservoirs and therefore that the detected CN is not a
photodissociation product of HCN. {This would imply that} CN emission
is not dominated by the upper, UV-exposed, disk layer, as is usually
assumed \citep{teague2016}. In contrast, CN emission from cold and
UV-shielded regions seems to be favoured. This is further supported by
the narrow CN linewidths which constrain the kinetic temperature to be
lower than 20--25~K within the CN ring and beyond (see
Section~\ref{app:excitation}). Observations of both CN and HCN at high
spatial resolution in a sample of disks spanning a range of viewing
geometries, coupled with comprehensive radiative transfer models, are
required to fully address the question of the specific origin of the
CN emission in disks.

\section{The \foun/\fifn\ isotopic ratio in the present-day ISM}

In the following, we provide arguments that the CN/C\fifn\ abundance
ratio we have directly measured in the disk orbiting TW~Hya is in fact
also a measurement of the elemental \nratio\ isotopic ratio in the
present-day LISM.

The CN abundance measured in the TW~Hya disk ratio agrees very well
with the average ratio in NH$_3$ of 321(36) towards Barnard~1, and in
N$_2$H$^+$ of 365(135) towards L1689E. Although ratios up to
$\sim1000$ have been obtained using N$_2$H$^+$ \citep{bizzocchi2013},
the weighted average of direct measurements performed in prestellar
cores is 336(16) (see Table~\ref{tab:ratios}) and is in very good
agreement with our CN ratio in TW~Hya. If one adopts the predictions
from selective photodissociation disk models \citep{heays2014}, CN in
TW~Hya would be enriched in $^{15}$N by $\approx80\%$ with respect to
the elemental reservoir. The elemental ratio in TW~Hya, and hence in
the LISM, would be $\rnow\approx 600$, thus larger than in the PSN 4.6
Gyr ago, at odds with galactic evolution chemical model
predictions. This would require ammonia and dyazelinium (\ce{N2H+}) in
dense cores to be similarly enriched by 80\%, in sharp disagreement
with predictions of chemical fractionation models
\citep{hilyblant2013b, roueff2015} (see
Section~\ref{app:fractionation}). This suggests that the efficiency of
selective photodissociation may be much lower in TW~Hya.

\begin{figure*}[t]
  \centering
  \includegraphics[width=\hsize]{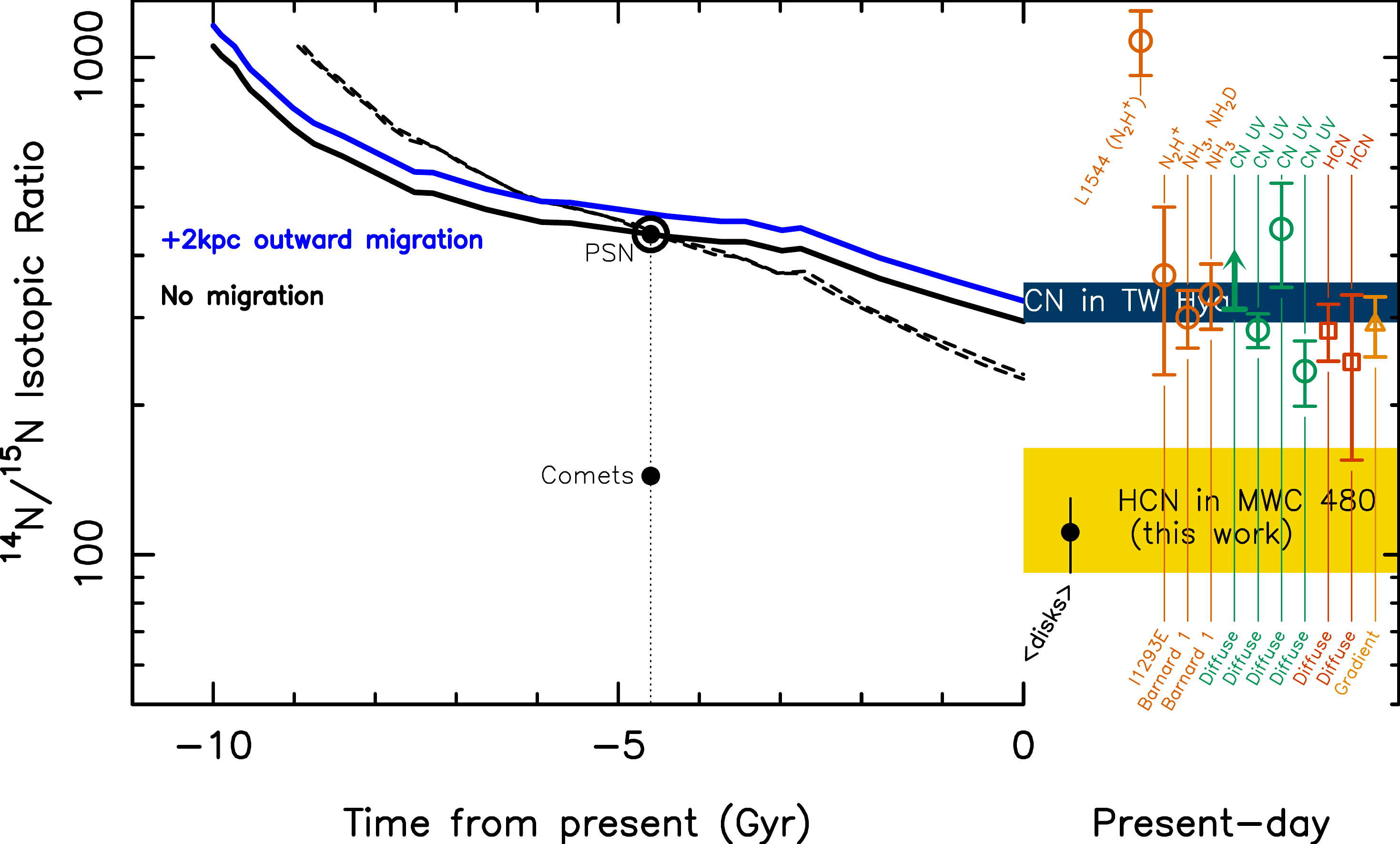}
  \caption{Present-day nitrogen isotopic ratio in the solar
    neighbourhood {compared to Galactic Chemical Evolution models
      predictions}. \emph{Shaded areas:} ALMA measurements of the
    CN/C$^{15}$N and HCN/HC\fifn\ ratios in TW~Hya and MWC~480
    respectively. \emph{Symbols}: directly measured ratios (i.e. not
    using double-isotopic ratios) in the present-day solar
    neighbourhood at $r_G=8.5(10)$~kpc (references are provided in
    Table~\ref{tab:ratios}). \emph{Black curves:} GCE model
    predictions \cite{romano2003} scaled to a PSN ratio of \rsun=441,
    with $^{15}$N as a primary (full line, their model 2) or secondary
    product (dashed lines, their models 3s an 3n). \emph{Blue curve:}
    \emph{GCE model 2} including a 2 kpc outward Sun migration (see
    the text for details). {The weighted average isotopic ratios in
      comets of 144(3) (see also Fig.~\ref{fig:comets}) and of
      111(19), from H\thcn/HC\fifn\ observations in 5 disks
      \citep{guzman2017}, are also indicated.}}
  \label{fig:gce}
\end{figure*}

Given the concomitant ratios in TW~Hya and in prestellar cores, a more
likely interpretation is that CN, NH$_3$, and N$_2$H$^+$, are tracing
a single reservoir of nitrogen. Furthermore, this reservoir would be
the primary one, therefore reflecting the elemental isotopic ratio in
the present-day solar neighbourhood. This scenario is in good
agreement with the HCN/HC$^{15}$N and CN/C$^{15}$N ratios in the
diffuse LISM (see Fig.~\ref{fig:gce} and
Table~\ref{tab:ratios}). Ratios derived from CN lines in the
ultraviolet (UV) show a large scatter, with ratios as high as 452(107)
, but yield an average ratio of 274(18)---although not compatible with
one lower limit at 312. Absorption spectroscopy using the (1-0)
rotational line of HCN in translucent clouds \citep[$A_V=1-4$ mag,
][]{gluck2016} yield a similar average of 276(34)
\citep{lucas1998}. In these clouds, CN is efficiently produced by
photodissociation of HCN, and both species most likely originate from
a single reservoir of nitrogen, presumably in atomic form
\citep{boger2005}. Chemical mass fractionation is inefficient in these
clouds ($T_k>15-30$~K), but selective photodissociation is predicted
to enhance $^{15}$N by 10\% in CN and HCN \citep{heays2014}. The low
ratios in diffuse clouds would thus correspond to an elemental ratio
of 304(37), in agreement with, although at the lower end of, the CN
ratio we have measured in TW~Hya.

{The conclusion that CN in TW~Hya traces the primary reservoir of
  nitrogen} also is in accord with predictions from galactic
models. According to purely dynamical models, the Sun most likely did
not migrate from its birthplace \citep{barbosa2015}, and the proposed
current ratio $\rnow=323(30)$ would indicate a $^{15}$N-enrichment by
up to 38(13)\% above \rsun\ over the last 4.6 Gyr. This {agrees,
  although marginally,} with GCE calculations including $^{15}$N
synthesis through novae outbursts (Fig.~\ref{fig:gce}), and especially
with models treating $^{15}$N as a primary product
\citep{romano2003}. {The agreement is much better with dynamical GCE
  models which instead predict that the Sun migrated outward by 0.3 to
  3.6 kpc \citep{minchev2013}. Indeed, applying the present-day
  galactic gradient $d\rr/dR=22.1$~kpc$^{-1}$ \citep{adande2012} to
  the Milky Way at 4.6~Gyr in the past, our CN/C\fifn\ ratio would
  correspond to an outward migration of the Sun by $\sim$2~kpc from
  its birth location, within the range predicted by dynamical GCE
  models. It is worth noting that, in addition to provide an
  explanation for the anomalously high solar metallicity with respect
  to young, local, B-type stars \citep{nieva2012}, the outward
  migration hypothesis would also bring the $^{12}$C/$^{13}$C
  elemental isotopic ratio from GCE models \citep{romano2003,
    milam2005} into a better agreement with the PSN value of
  $\approx90$.}

\section{The origin of nitrogen in the solar system}

{Our inference that a fractionated reservoir of nitrogen is present at
  the time of comet formation in PSN analogues sheds new light on the
  history of nitrogen in the solar system. The presence of such a
  fractionated nitrogen reservoir in the primitive solar system would
  indicate that} parent body evolution is not required to explain the
average cometary ratio of 144(3) (see Fig.~\ref{fig:comets}). A more
likely explanation is that atomic nitrogen, from which HCN presumably
derives \citep{hilyblant2013a}, is a secondary repository of nitrogen
at the PSN stage, supporting the usual assumption that N$_2$
constitutes the primary reservoir \citep{schwarz2014, furi2015}. In
this interpretation, the isotopic ratio of 140 in NH$_3$ in comets
\citep{shinnaka2016b} could indicate that cometary ammonia was
primarily formed by hydrogenation of atomic nitrogen in CO-rich
{interstellar} ices \citep{fedoseev2015a}. In addition, the ring-like
distribution of CN in TW~Hya, which encompasses the giant planet and
Kuiper-belt regions, suggests that the primary reservoir was present
in the comet-forming zone. Yet, the uniform isotopic ratio of 144 in
all comets and the low N$_2$/CO abundance ratio in comets Halley and
67P/C-G \citep{rubin2015a} strongly suggest that the primary reservoir
was lost, or not captured, by comets, although the underlying
mechanism remains an open question. An alternative would be that the
bulk of nitrogen in comets has still escaped detection.  It is also
striking that the \emph{ratio} of the CN- and HCN-derived nitrogen
isotopic ratios in disks, 323:111 ($\approx$2.9), are in the same
proportion as the mean PSN ratio, 441:144 ($\approx$3.1), as measured
in the Sun and comets. This further suggests that the fractionation
processes building the two isotopic reservoirs of nitrogen, evinced in
present-day PSN analogues, were equally active in the PSN, or in the
prestellar core where it formed, 4.6 Gyr ago.

A picture emerges in which dinitrogen was the primary reservoir of
nitrogen in the PSN, with an isotopic ratio of 441, whereas atomic N
formed a secondary reservoir characterized by \rr$\approx$140. Earth's
atmosphere appears intermediate (\rr=272), {which could reflect} a
mixture of the two reservoirs of volatile nitrogen, {or a third
  reservoir in its own right \citep{furi2015}}. The remarkably uniform
ratio in comets {(see Fig.~\ref{fig:comets})} suggests that both
reservoirs were homogeneously distributed in the PSN. Disentangling
the origin of the fractionated reservoir, between processes in the PSN
and an interstellar origin, as suggested by the HCN isotopic ratios in
prestellar cores \citep{hilyblant2013a}, requires high-sensitivity
maps of the isotopic ratio in cores and especially of disks spanning a
broad range of physical conditions. {However, the small scatter of the
  HCN/HC\fifn\ ratios indirectly measured towards a sample of disks
  \citep{guzman2017} spanning a broad range of ages and masses---hence
  likely representing a range of dust size distributions and UV
  radiation fields---argues in favour of a prestellar
  origin. Moreover, HCN/HC\fifn\ isotopic ratios as low as 140 were
  indirectly obtained in prestellar cores
  \citep{hilyblant2013a}. Nevertheless, direct measurements of
  HCN/HC\fifn\ in prestellar cores remain the missing clue to put the
  interstellar origin of the fractionated reservoir on a firmer
  ground. If the scenario presented here holds, we would predict that
  the HCN/HC\fifn\ ratio in prestellar cores should indeed be lower
  than the cometary ratio of 144, and closer to the average ratio in
  disks of 111(19), as a result of galactic chemical evolution.}

\section{Conclusions}

Measurements of nitrogen isotopic ratios are keys to understanding the
origin and evolution of the solar system in the context of Galactic
nucleosynthesis. The large range of measured \foun/\fifn\ isotopic
ratios in solar system bodies, from $\approx$440 in the Sun and
Jupiter to $\approx$140 in comets, may reflect the presence of
distinct reservoirs of nitrogen that have recorded interstellar
initial conditions, chemical fractionation processes in the PSN,
and/or secular elemental abundance evolution over the past 4.6 billion
years. In this paper, we have presented the first direct measurement
of the volatile C\foun/C\fifn\ isotopic ratio within a protoplanetary
disk. The ratio of 323(30), obtained with the ALMA interferometer, is
in good agreement with the ratio measured for prestellar cores. The
CN/C\fifn\ abundance ratio in the TW~Hya disk therefore likely
reflects a non-fractionated reservoir of nitrogen that is
representative of the present-day elemental ratio of nitrogen in the
solar neighbourhood. {The comparison with the indirectly
  derived HCN/HC\fifn\ ratio in other disks demonstrate that disks at
  the comet formation stage already contain a heavily fractionated
  reservoir. The measurement of the HCN/HC\fifn\ ratio in TW~Hya would
  provide a striking confirmation of this finding.} On a broader
perspective, our results support galactic chemical evolution models
invoking novae outbursts as primary \fifn\ sources as well as
significant outward migration of the Sun over its lifetime, and
furthermore suggest that, whereas the Sun and Jupiter incorporated
mass from the PSN's main nitrogen reservoir, solar system comets
recorded a secondary, fractionated reservoir of nitrogen.

\bibliography{general,allbooks}
\bibliographystyle{aa}

\begin{acknowledgements}
  PHB acknowledges the \emph{Institut Universitaire de France} for
  financial support. PHB also thanks L. Bonal, N. Prantzos, and
  E. Quirico for fruitful discussions. The authors wish to thank the
  referee for a careful reading and for useful comments which improved
  the clarity of the paper. This paper makes use of the CDMS catalogue,
  the CASA and IRAM/GILDAS softwares, and of ALMA data. ALMA is a
  partnership of ESO (representing its member states), NSF (USA) and
  NINS (Japan), together with NRC (Canada), NSC and ASIAA (Taiwan),
  and KASI (Republic of Korea), in cooperation with the Republic of
  Chile. The Joint ALMA Observatory is operated by ESO, AUI/NRAO and
  NAOJ. This work has made use of data from the European Space Agency
  (ESA) mission {\it Gaia} (\url{https://www.cosmos.esa.int/gaia}),
  processed by the {\it Gaia} Data Processing and Analysis Consortium
  (DPAC,
  \url{https://www.cosmos.esa.int/web/gaia/dpac/consortium}). Funding
  for the DPAC has been provided by national institutions, in
  particular the institutions participating in the {\it Gaia}
  Multilateral Agreement.
\end{acknowledgements}

%newpage
\begin{appendix}
\clearpage
\section{Supplementary materials}

\begin{table*}
  \begin{center}
  \caption{\label{tab:spectro} Spectroscopic properties and relative
    intensities (R.I.) of the hyperfine CN and C$^{15}$N $(NJF\rightarrow
    N'J'F')$ transitions. }
  \begin{tabular}{llllcccc}
    \hline
    Species & Group & $N J F^\S$     & $N' J' F'$ & Rest frequency & $A_{ul}^\dag$
    & $g_u^\ddag$ & R.I.$^\sharp$ \\
    & & & & (MHz) &  (s$^{-1}$)\\
    \hline
    CN
    & b
    &  3 5/2 5/2 & 2 3/2 5/2 & 340008.13 & 6.197$\times 10^{-5}$ & 6 & 0.054\\
    && 3 5/2 3/2 & 2 3/2 3/2 & 340019.63 & 9.270$\times 10^{-5}$ & 4 & 0.054\\
    && 3 5/2 7/2 & 2 3/2 5/2 & 340031.55 & 3.845$\times 10^{-4}$ & 8 & 0.445\\
    && 3 5/2 3/2 & 2 3/2 1/2 & 340035.41 & 2.887$\times 10^{-4}$ & 4 & 0.167\\
    && 3 5/2 5/2 & 2 3/2 3/2 & 340035.41 & 3.231$\times 10^{-4}$ & 6 & 0.281\\
    & a
    &  3 7/2 7/2 & 2 5/2 5/2 & 340247.77 & 3.797$\times 10^{-4}$ & 8 & 0.307\\
    && 3 7/2 9/2 & 2 5/2 7/2 & 340247.77 & 4.131$\times 10^{-4}$ &10 & 0.417\\
    && 3 7/2 5/2 & 2 5/2 3/2 & 340248.54 & 3.674$\times 10^{-4}$ & 6 & 0.222\\
    && 3 7/2 5/2 & 2 5/2 5/2 & 340261.77 & 4.479$\times 10^{-5}$ & 6 & 0.027\\
    && 3 7/2 7/2 & 2 5/2 7/2 & 340264.95 & 3.350$\times 10^{-5}$ & 8 & 0.027\\
    C$^{15}$N                                                       
    && 3 5/2   2 & 2 3/2   1 & 329623.32 & 3.155$\times 10^{-4}$ & 5 & 0.159\\
    && 3 5/2   3 & 2 3/2   2 & 329625.98 & 3.508$\times 10^{-4}$ & 7 & 0.247\\
    && 3 7/2   3 & 2 5/2   2 & 329837.13 & 3.583$\times 10^{-4}$ & 7 & 0.253\\
    && 3 7/2   4 & 2 5/2   3 & 329837.65 & 3.764$\times 10^{-4}$ & 9 & 0.341\\
    \hline
  \end{tabular}
\end{center} {\footnotesize {Notes:} $^\S$ Quantum numbers:
  ${\mathbf N}$ is the rotational angular momentum, $\mathbf{J=N+1/2}$
  describes the electronic spin coupling, and $\mathbf{F=J+I}$
  describes the nuclear spin coupling {($I=1$ for \foun, and
    1/2 for \fifn)}.  $^\dag$ Einstein coefficient for spontaneous
  radiative decay.  $^\ddag$ Upper level degeneracy (including fine
  and hyperfine splitting).  $^\sharp$ Normalized to a total of unity
  per fine structure group.}
  % \begin{list}{}{}
  %   \footnotesize
  % \item[$\S$] Quantum numbers: ${\mathbf N}$ is the rotational angular
  %   momentum, $\mathbf{J=N+1/2}$ describes the electronic spin
  %   coupling, and $\mathbf{F=J+1/2}$ describes the N atom nuclear
  %   spin.
  %   \item[$\dag$] Einstein coefficient for spontaneous radiative decay.
  %   \item[$\ddag$] Upper level degeneracy (including fine and hyperfine splitting).
  %   \item[$\sharp$] Normalized to a total of unity per fine structure
  %     group.
  % \end{list}
\end{table*}

\begin{figure*}
  \centering
  \includegraphics[width=.9\hsize]{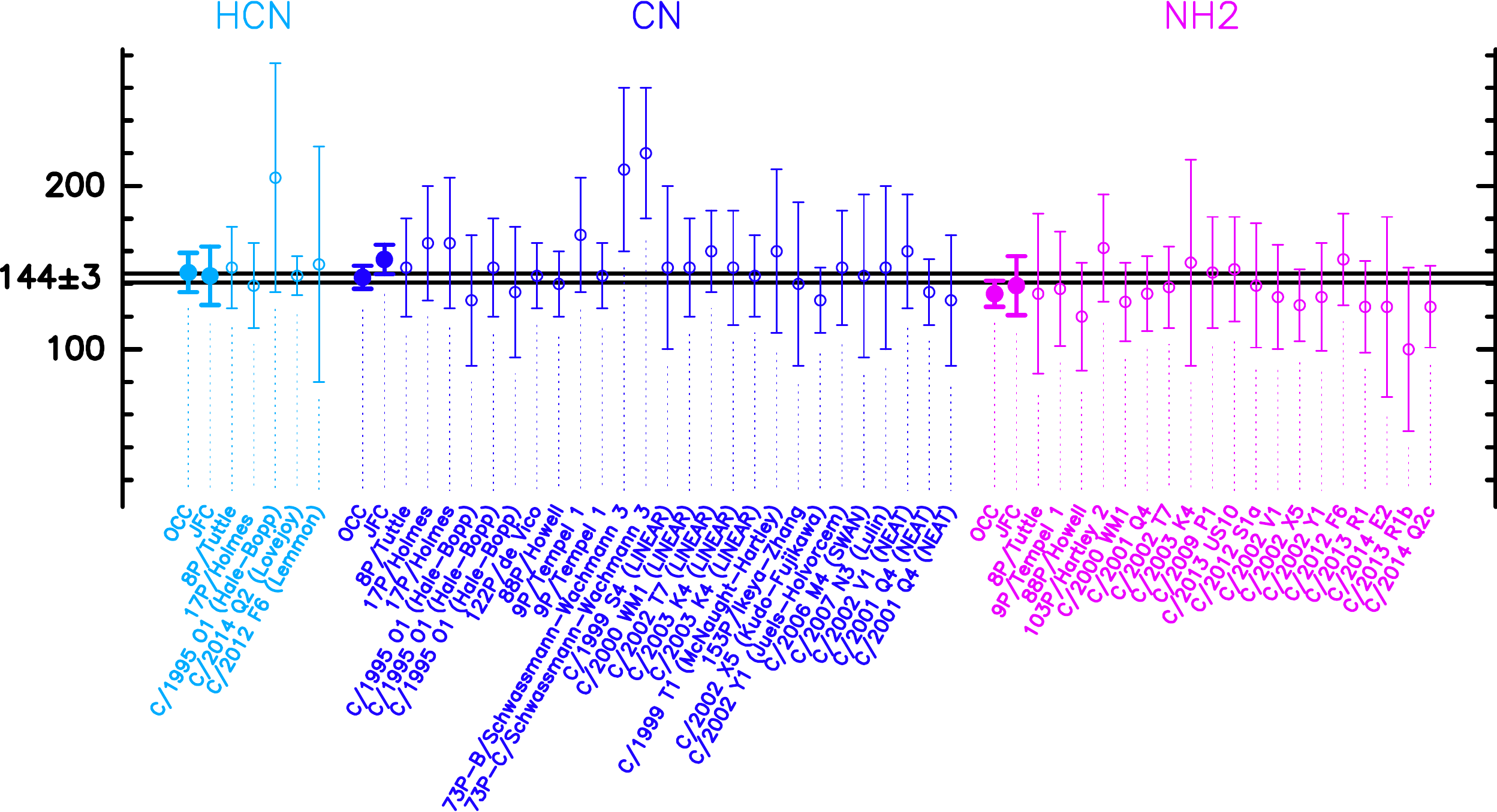}
  \caption{{Compilation of all isotopic ratios in comets. The observed
      carrier is indicated above each measurement. Although it is
      clear that \ce{NH2} is a photodissociation product of ammonia in
      the coma, the origin of CN is less certain. Data were taken from
      \cite{jehin2009} and \cite{bockelee2015}. The weighted average
      ratio is 144(3). For each carrier, the average ratios for
      Jupiter Family Comets (JFC) and Oort Cloud Comets (OCC), are
      indicated, where the distinction between JFC and OCC was based
      only on the periodicity of the comets being respectively lower
      and higher than 200~yr \citep{shinnaka2016b}.}}
  \label{fig:comets}
\end{figure*}

\clearpage
\newpage

% \begin{figure}
%   \centering
%   \includegraphics[width=.9\hsize]{c14n-340020-cmap}
%   \caption{Channel maps of the optically thin hyperfine lines of CN,
%     at 340.020~GHz (in Jy/beam). The LSR velocity is indicated (in
%     km/s) in each panel. Offsets are in arcsec, with respect to the
%     phase center. The half-power beamwidth is $0.89"\times0.53"$.}
%   \label{fig:cmapa}
% \end{figure}
% \begin{figure}
%   \centering
%   \includegraphics[width=.9\hsize]{c14n-340265-cmap}
%   \caption{Same as Fig.~\ref{fig:cmapa}, for the optically thin
%     hyperfine component at at 340.265~GHz.}
%   \label{fig:cmapb}
% \end{figure}

\begin{figure*}
  \centering
  \includegraphics[width=.45\hsize]{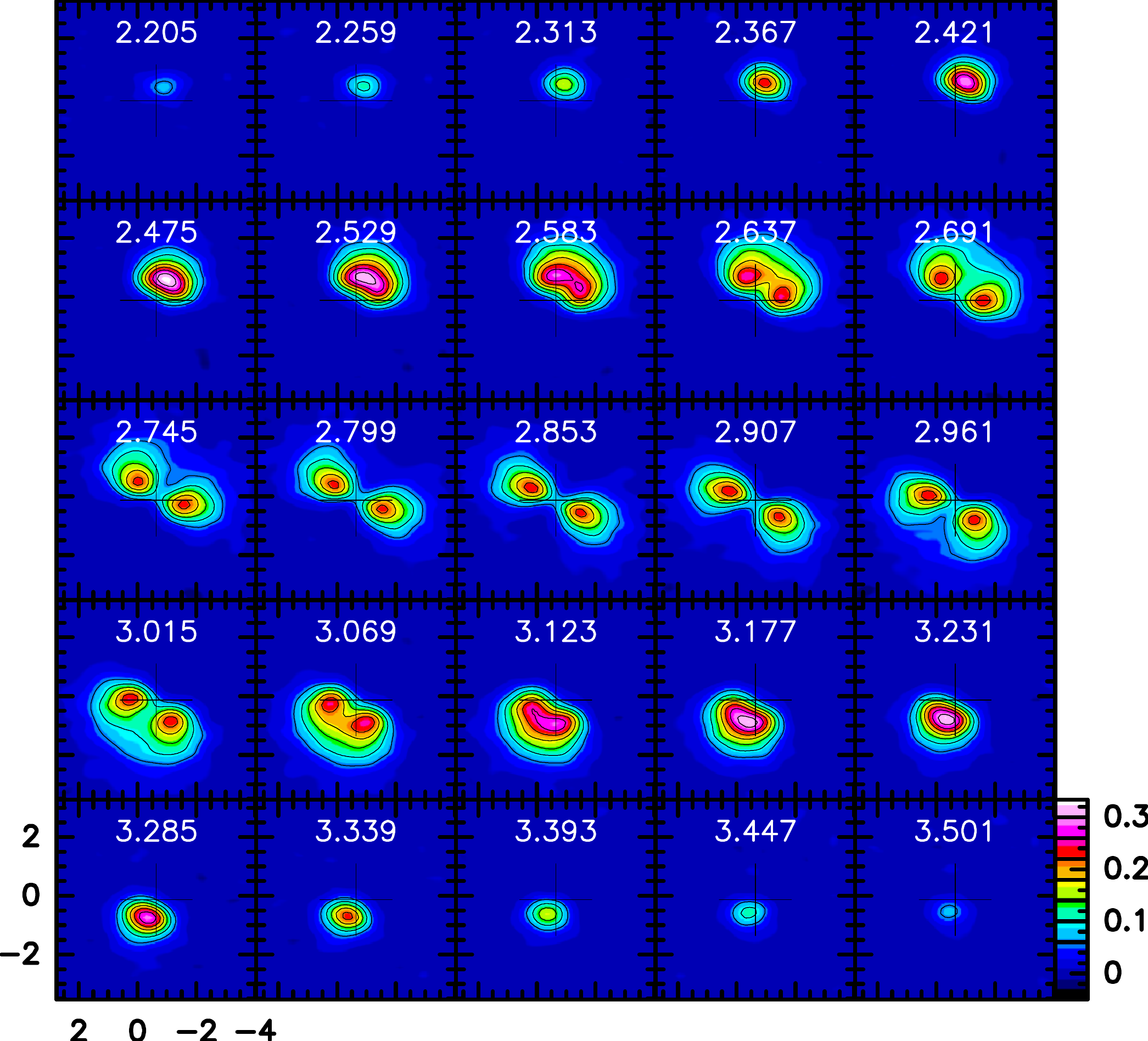}\hfill%
  \includegraphics[width=.45\hsize]{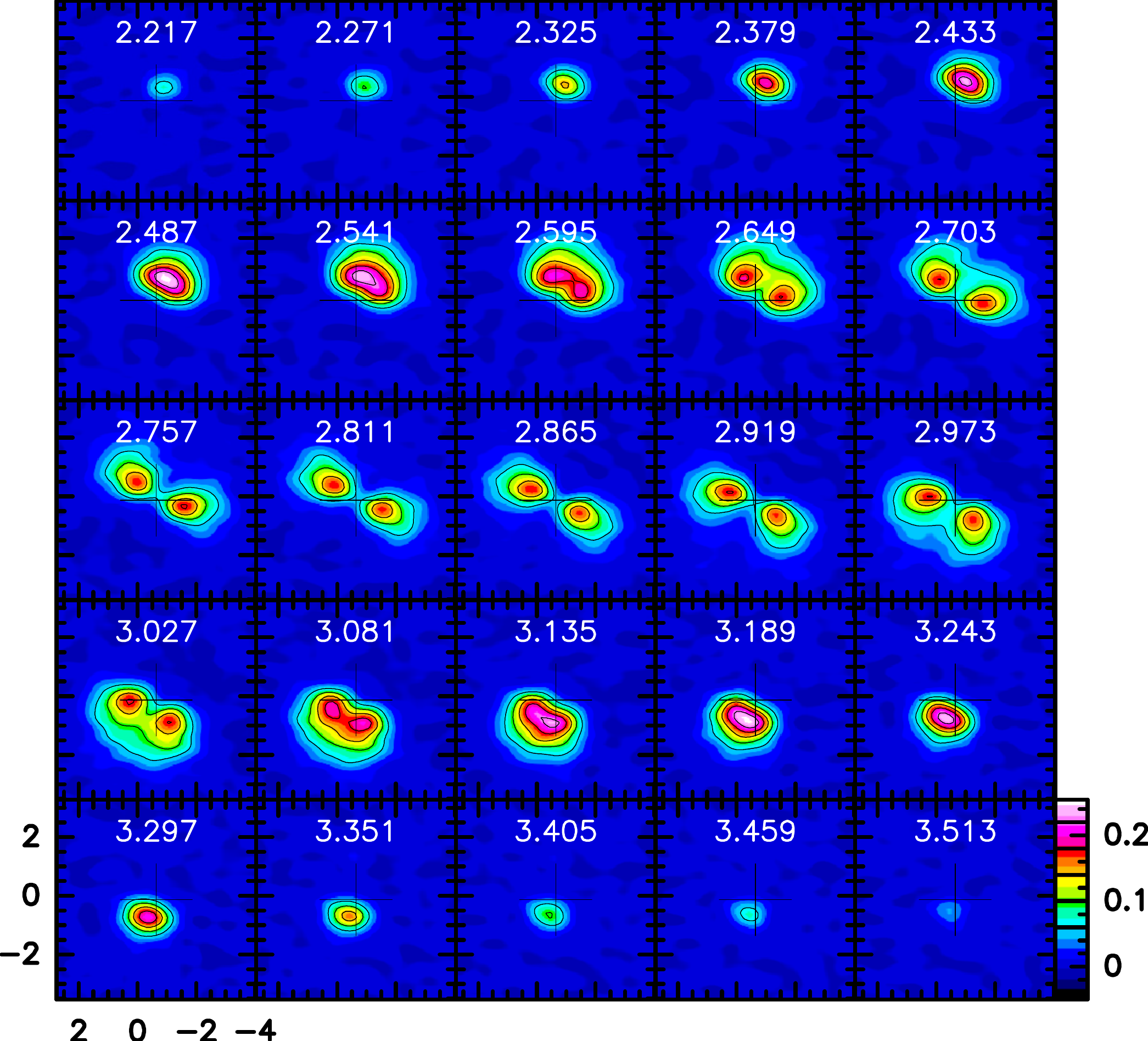}
  \caption{Channel maps (in Jy/beam) of the optically thin hyperfine
    lines of CN at 340.020~GHz (left) and 340.265~GHz (right). The LSR
    velocity is indicated (in km/s) in each panel. Offsets are in
    arcsec, with respect to the phase center. The half-power beamwidth
    is $0.89"\times0.53"$.}
  \label{fig:cmaps}
\end{figure*}

\clearpage
\newpage

% We improved the signal-to-noise (SNR) on the C$^{15}$N dataset by
% tapering the visibilities resulting in maximum baselines of 120~m and
% leading to a synthesized beam of 2.3\arcsec.  The untapered CN dataset
% has a synthesized beam of 0.89$\times$0.53\arcsec\ and a position
% angle of 86~\degr. The final sensitivity is between 1.8 and
% 3.6~mJy/beam/channel for the CN(3-2) spectra, and between 5.3 to 6.4
% mJy/beam/channel in C$^{15}$N.

\section{$uv$-plane analysis}
\label{app:analysis}

\begin{figure*}
  \centering
  \includegraphics[width=0.8\hsize]{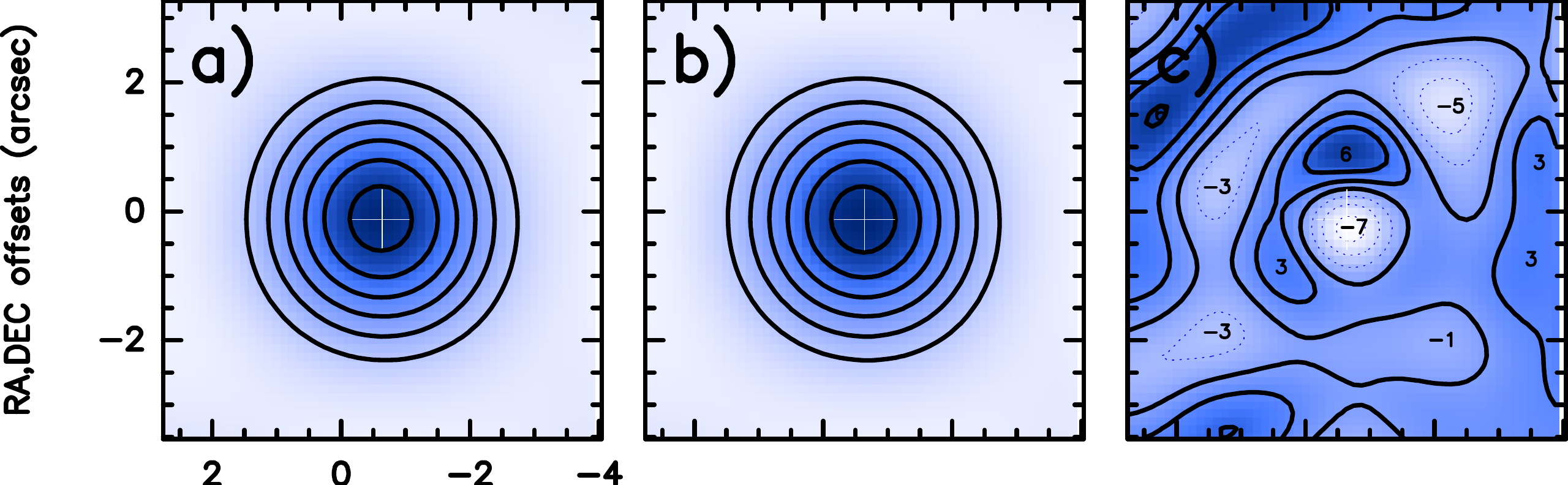}\bigskip\\
  \includegraphics[width=0.8\hsize]{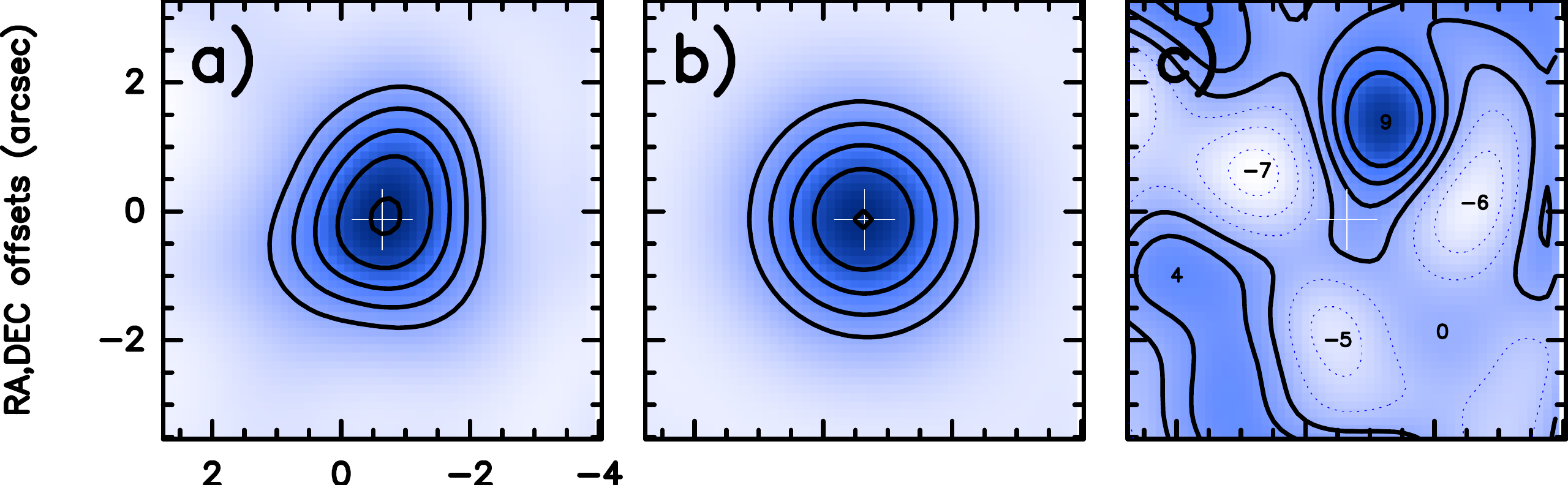}
  \caption{Results of the $uv$-plane analysis (see the text for
    details) for C$^{14}$N\ (top) and C$^{15}$N (bottom). On each row,
    panels a) an b) show the dirty image and the fitted Gaussian disk
    (Jy/beam), and panel c) shows the residuals (in mJy/beam). For
    C$^{14}$N, contours in panels a) and b) are 0.1 to 0.7 by 0.1
    Jy/beam, and for C$^{15}$N from 0.02 to 0.07 by 0.01 Jy/beam. Values
    of local extrema are indicated in the residuals. The C$^{15}$N fit
    assumes the Gaussian disk from C$^{14}$N. The results are summarized
    in Table~\ref{tab:uvfit}.}
  % \includegraphics[width=\hsize]{1GAUSS_cnb_uv60}\bigskip\\
  % \includegraphics[width=\hsize]{1GAUSS_c15n_uv60_circular}\\
  % \caption{Results of the $uv$-plane analysis (see the text for
  %   details). The intensity scale is proportional to the column
  %   density, and the C$^{15}$N scale is normalized to a CN:C$^{15}$N ratio
  %   of 1:441. On each row, from left to right: a) is the dirty image
  %   tapered to $uv$ radii $\le$60~m, b) is the fitted 2-D Gaussian,
  %   and c) are the residuals (in mJy/dirty-beam). In our re-scaled
  %   units, peak values are 2.2 and 3.3 Jy/dirty-beam for CN and
  %   C$^{15}$N respectively. Contours are from 0.2 to 2.2 by 0.4
  %   Jy/dirty-beam in a) and b) panels and, in the residuals maps, from
  %   -20 to 20 by 10 for CN and -400 to 400 by 200 (mJy/dirty-beam) for
  %   C$^{15}$N. The C$^{15}$N results are from a circular Gaussian (see
  %   Table~\ref{tab:uvfit}).}
  \label{fig:uvfit}
\end{figure*}

The integrated visibilities of CN and C$^{15}$N were fitted with
two-dimensional Gaussian while applying a cutoff radius (equivalent to
smoothing in the direct plane), from 60 to 120~m. In particular, with
cutoffs smaller than 100~m, the central hole is smeared out and the
Gaussian-disk model is mechanically appropriate.  The fitting was
performed using the \verb+UV_FIT+ task of the \textsc{Gildas/Mapping}
software. The 6 free parameters of the CN fit are the center, major
and minor axis, the position angle of the Gaussian disk, and the total
flux. The geometrical parameters of the CN disk are applied to the
C$^{15}$N visibilities thus leaving the flux as the only one free
parameter. The results depend very marginally on the cutoff radius,
although the residuals of the CN fits obviously show an increasing
contribution from the central hole as the spatial resolution
improves. The results of the fitting procedure for a cutoff $uv$
radius of 60~m are shown in Fig.~\ref{fig:uvfit} and summarized in
Table~\ref{tab:uvfit}.

To validate the procedure, we also performed a Gaussian fit to the
C$^{15}$N visibilities while letting all parameters as free, leading
to geometrical parameters {which agree} with the CN fit to within
1$\sigma$ or less (see Table~\ref{tab:uvfit}). To estimate the
contribution of the geometrical assumptions to the final uncertainties
of the fluxes and flux ratio, we performed a series of $uv$
minimizations, in which the same (either elliptical or circular)
Gaussian shape was imposed to both CN and C$^{15}$N, with the center
and size parameters randomly and independently chosen within 2$\sigma$
around the previous CN fit result. The resulting fluxes and flux
ratios vary by typically 0.1$\sigma$, showing that the uncertainty is
dominated by the 10\% statistical fluctuations of the C$^{15}$N
visibilities.

Of the above assumptions, the co-spatial distribution of CN and
C$^{15}$N is likely the most difficult to firmly assess, as it ideally
rests on self-consistent physico-chemical models of disks, in a
time-dependent fashion to properly take into account the complex
interplay between physical and chemical processes acting on comparable
timescales.  Moreover, radiative transfer effects can lead to
different distribution-to-emission relation even for co-spatial
species. However, having optically thin lines in hand greatly
alleviates such issues. Still, even if the main production and
destruction pathways of isotopologues are the same, specific reactions
may proceed at different rates, giving rise to fractionation. The
co-spatial assumption is therefore species- and
environment-dependent. In the case of CN in TW~Hya, the potential
fractionation processes are chemical reactions and selective
photodissociation. Chemical fractionation is most likely negligible
for CN, because of inefficient fractionation reactions
\citep{roueff2015} and/or because freeze-out is damping out gas-phase
processes \citep{heays2014}. Selective photodissociation could be a
radially variable process, since it is driven by UV photons while
being expectedly sensitive to the dust size distribution, two
properties which depend on the distance to the central protostar and
height above the midplane. The outcome of both effects is far from
straightforward, and could even cancel each other, with for example
dust growth at large radii compensating for geometrical dilution of
the UV photons from the central protostar (if dominating the UV
flux).% Nevertheless, in the present analysis, the radially smoothed CN
% and C$^{15}$N emissions are assumed to emanate from identical Gaussian
% disk, regardless of their precise spatial distribution accross the
% disk. The influence of this assumption was tentatively evaluated by
% performing a $uv$ fit to the C$^{15}$N data allowing for free center
% and size of an elliptical Gaussian, leading to consistent results.
We note that the maps in Fig.~\ref{fig:maps} suggest that the
C$^{15}$N integrated intensity is shifted Nort-West with respect to
the CN map. However, we stress that these maps require deconvolving
the visibilities which always introduces artifacts, especially when
the S/N is not high, as is the case for C$^{15}$N, and no strong
effort (apart from using robust weighting) was made to try and improve
its deconvolution. Indeed, the C$^{15}$N map was not used in this
work.

\section{Excitation of CN and C$^{15}$N}
\label{app:excitation}

One key feature of our derivation of the CN/C$^{15}$N abundance ratio
is the detection, with high S/N, of the two weakest hyperfine
transitions of CN, at 340.262 and 340.265~GHz, thus providing direct
access to its total column density with no requirement for radiative
transfer calculations. Being able to measure the CN/C$^{15}$N
abundance ratio directly from the main isotopologue also improves
significantly over previous observations in prestellar cores using the
C$^{13}$N/C$^{15}$N double isotopic ratio \citep{hilyblant2013b}.

\begin{figure*}[t]
  \centering
  \includegraphics[height=.4\textheight]{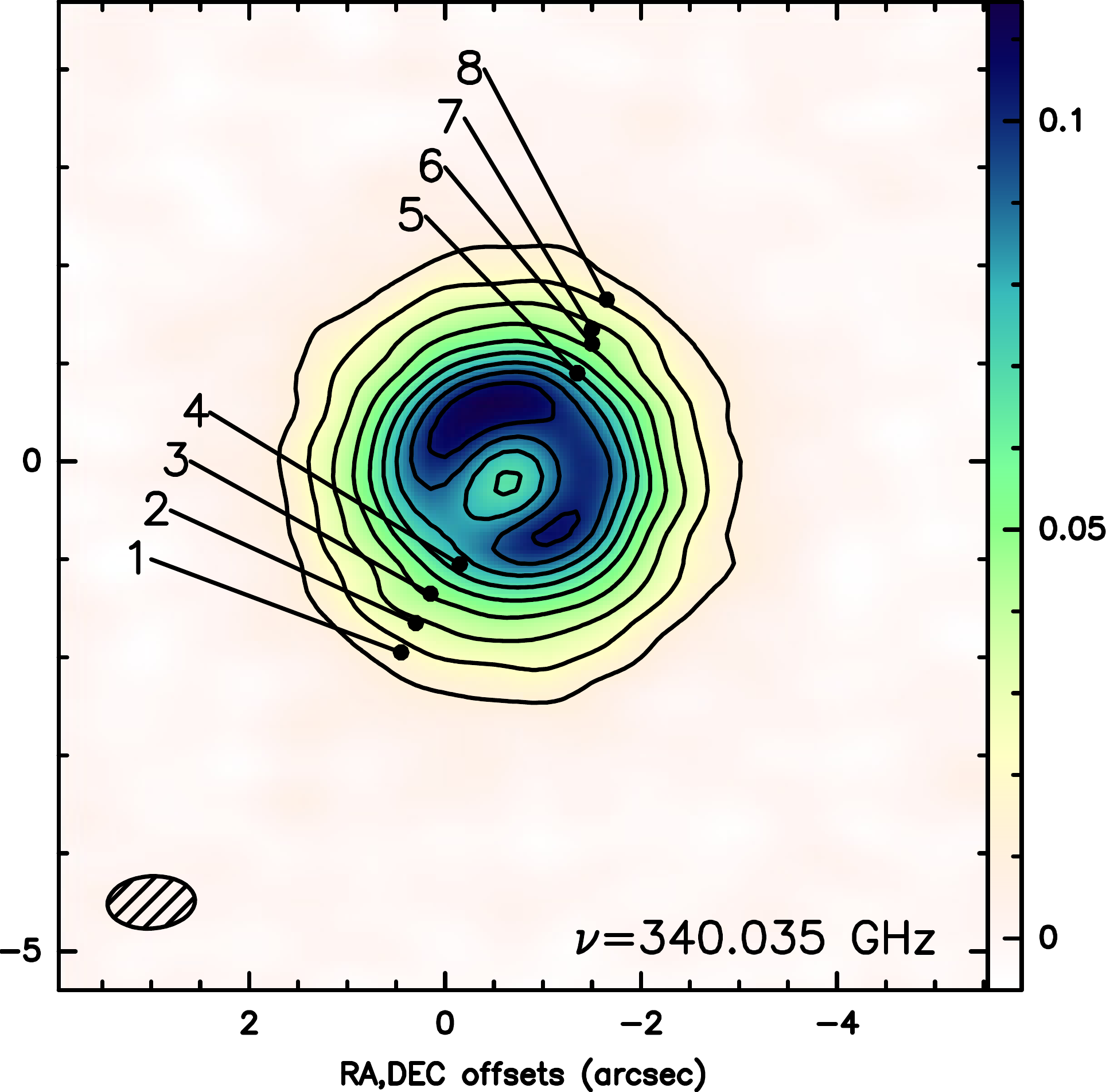}\bigskip\\
  \includegraphics[width=0.45\hsize]{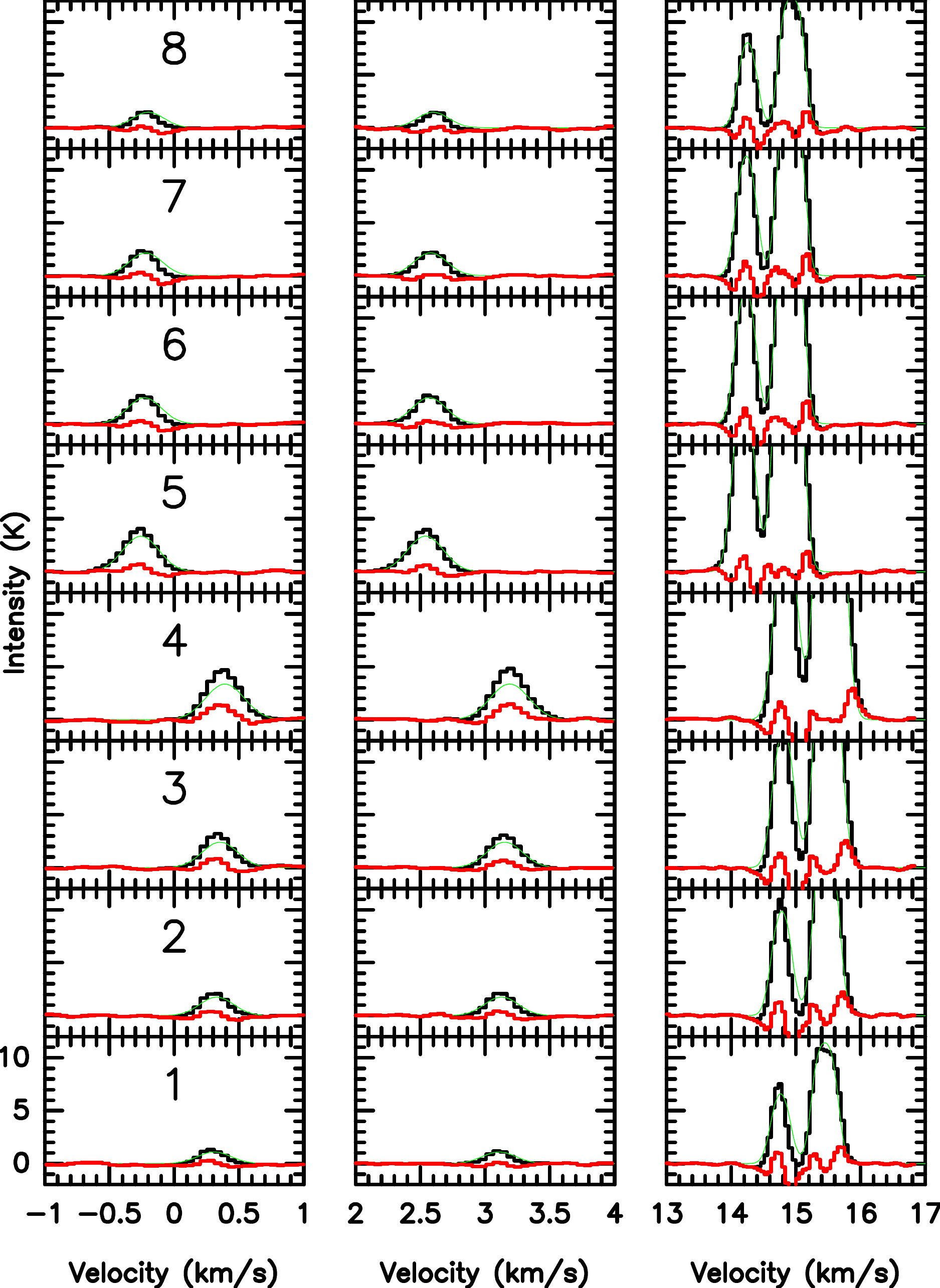}\hfill%
  \includegraphics[width=0.45\hsize]{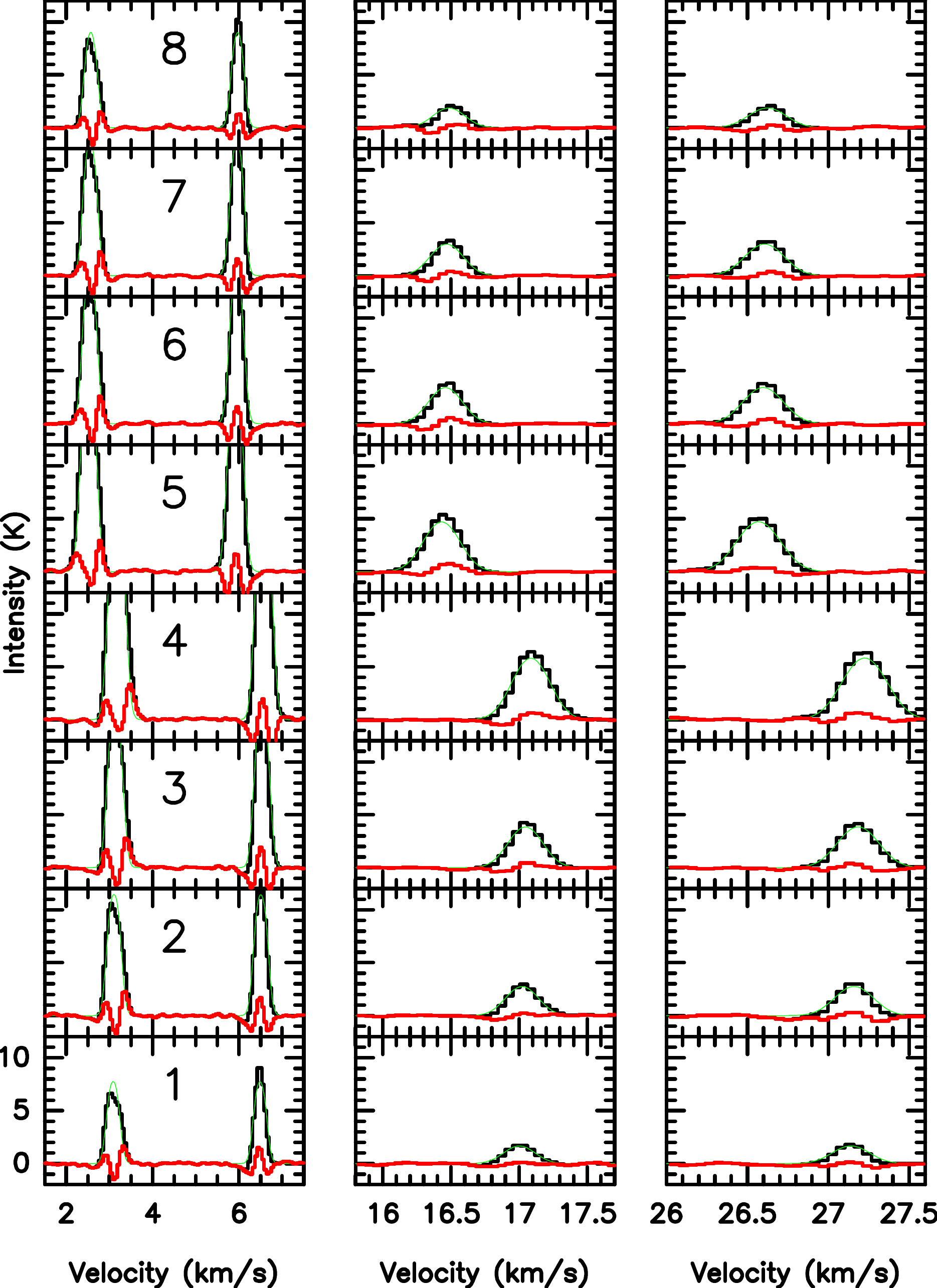}
  \caption{{Spectra of the hf lines of CN ($a$ group, left, $b$ group
      right) at different locations in the disk (top panel; integrated
      emission, in Jy/beam). Fits and their residuals to hyperfine
      structure (see Sec.~\ref{app:excitation} for details) are shown
      for each hf transition (green and red, resp.). Only reduced
      velocity intervals centered on the hf lines are shown.}}
  \label{fig:hfsa}
\end{figure*}

We estimated the optical depth of the hf components of the CN
rotational transitions by extracting deconvolved CN spectra at
different locations accross the disk (see Fig.~\ref{fig:hfsa}). In
what follows, the hf lines of the fine structure group at rest
frequencies around 340.3~GHz are labelled as $a$, while $b$ refers to
those around 340.0~GHz (see Table~\ref{tab:spectro}). At each
position, each group of hf lines was fitted simultaneously, using the
\verb+HFS+ fitting procedure \citep{hilyblant2013a} of the
\textsc{Gildas/Class} software. In doing so, the fluxes were brought
into the specific-intensity scale by applying a 44~mJy/K conversion
factor. The fitting algorithm assumes that the excitation temperature
is equal for all hf lines within each group (see below). The results
are shown in Fig.~\ref{fig:hfsa}. Towards the central parts of the
disk, where Keplerian smearing increases, the minimization becomes
poorer than towards the external parts where the lines are very well
fitted. Fitting each fine structure set of lines provides a
consistency check of the results.

\begin{figure}
  \centering
  \includegraphics[width=0.8\hsize]{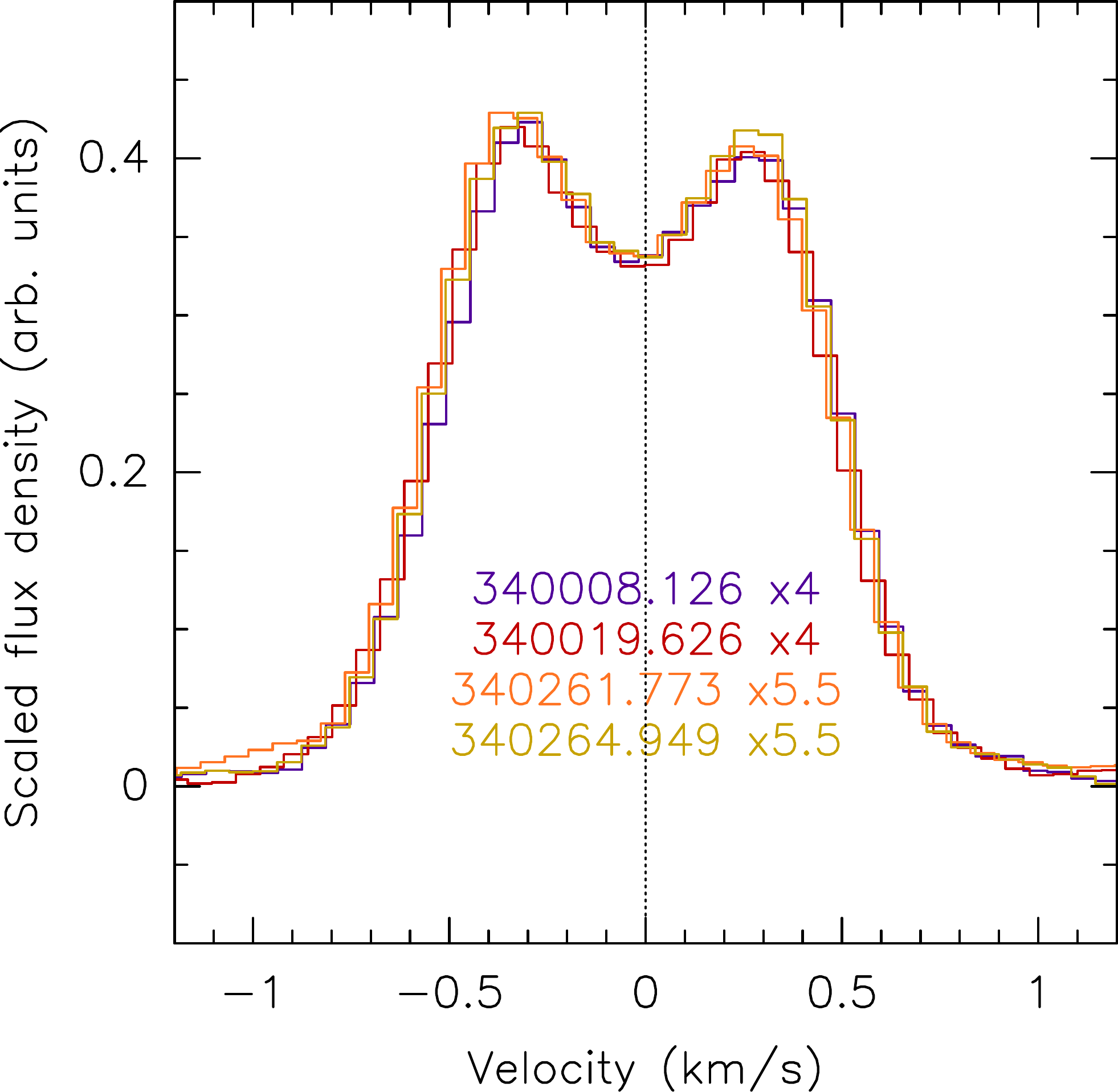}
  \caption{Disk-averaged four weakest hf transitions of CN(3-2) (see
    Table~\ref{tab:spectro}) are superimposed after being scaled by
    their theoretical relative intensities.}
  \label{fig:stacking}
\end{figure}

The derived total opacity is $\approx 3.4-5.4$ for groups $a$ and $b$
lines, confirming previous studies \citep{kastner2014}. The weakest hf
CN lines, in group $a$, which carry 2.7\% of the total intensity (see
Table~1), is thus less than 0.15, well into the optically thin
regime. {Figure~\ref{fig:stacking} shows a comparison of the
  disk-averaged spectra of the four weakest hf transitions of CN
  scaled by their relative intensities. The very good match of the
  spectral profiles indicates that they are all optically thin, as no
  opacity broadening can be noticed although the relative intensities
  vary from 2.7 to 5.4\%.}  The detected C$^{15}$N hf lines can safely
be assumed optically thin. This is further supported by the observed
relative strengths of these two hf lines (see Fig.~\ref{fig:spectra}).

% \begin{figure}
%   \centering
%   \includegraphics[height=0.33\textheight]{width-radius-allhf}
%   \caption{Azimuthal average of the FWHM (in km/s) of hf lines of
%     CN(3-2). The corresponding upper limit on the kinetic temperature}
%   \label{fig:ringfwhm}
% \end{figure}
\clearpage
\newpage

% \begin{figure}[t]
%   \centering
%   \includegraphics[width=\hsize]{mkplot-cna}
%   \caption{Same as Fig.~\ref{fig:hfsa} for the $b$ group of hf CN
%     transitions.}
%   \label{fig:hfsb}
% \end{figure}

\clearpage

\begin{figure}
  \centering
  \includegraphics[width=0.8\hsize]{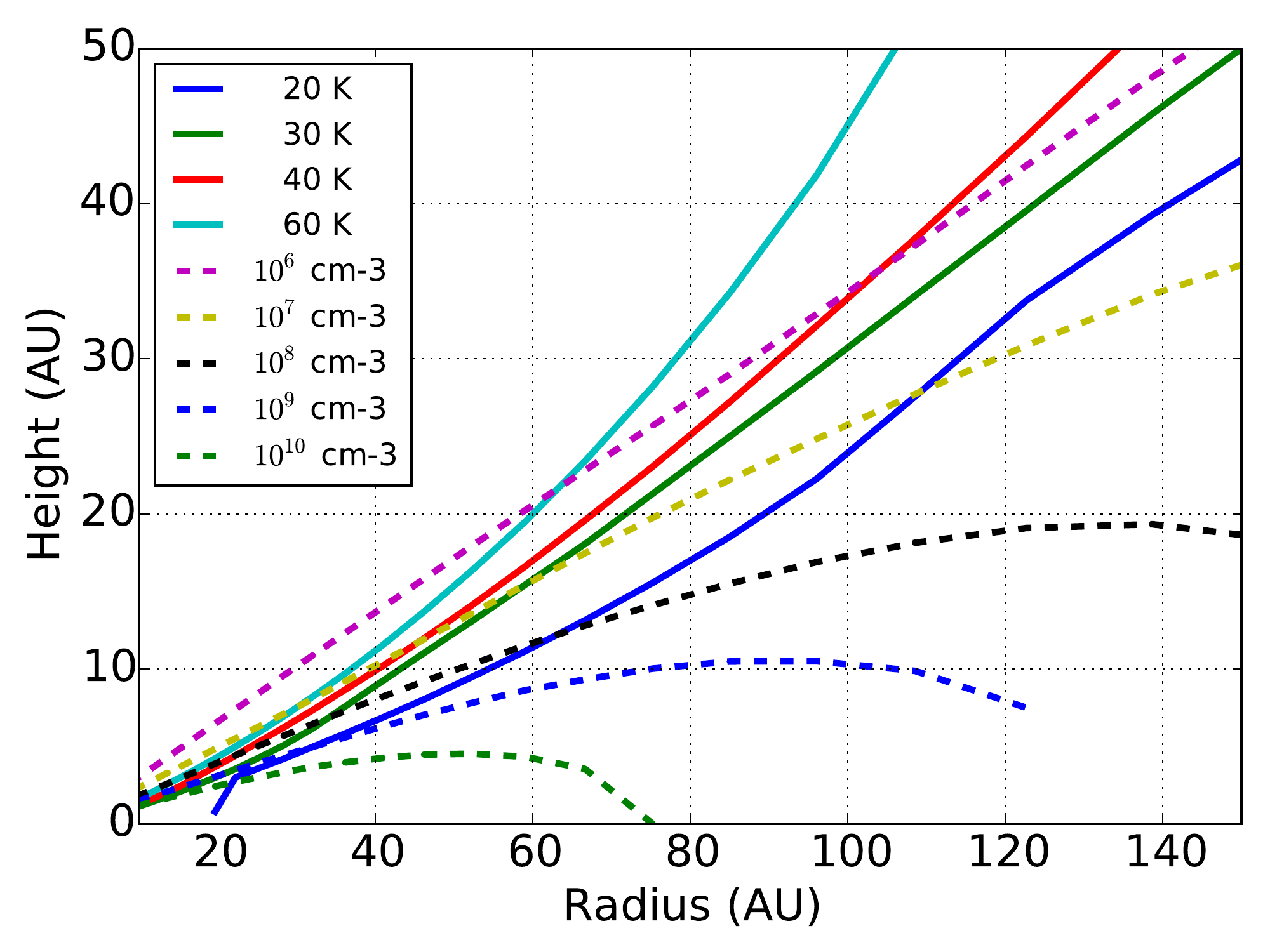}
  \caption{Locus of constant kinetic temperature (in K) and total
    number density (in cm$^{-3}$) from the physical TW~Hya model of
    \citep{qi2013}.}
  \label{fig:isolines}
\end{figure}

For optically thin lines, going from line intensity ratios to
abundance ratios only requires the spectroscopic properties and
excitation of each line to be taken into account, based on the usual
relation between the total column density and the integrated opacity:
\begin{equation}
  \label{eq:cdens}
  N_{\rm tot} = \frac{8\pi\nu^3}{A_{ul} g_uc^3}
  \frac{e^{E_l/kT_{\rm ex}} Q(T_{\rm ex})}{1-e^{-(E_u-E_l)/kT_{\rm ex}}}
  \,\int \tau_\nu \,{\rm d} \nu = f(ul,T_{\rm ex})\,\int \tau_\nu \,{\rm d} \nu,
\end{equation}
where $T_{\rm ex}$ is the excitation temperature of the
$u\rightarrow l$ hyperfine transition characterized by its Einstein
coefficient $A_{ul}$ and upper level degeneracy $g_u$ (see
Table~\ref{tab:spectro}). The partition function $Q$ is also a
function of the excitation temperature, and is calculated through a
linear interpolation accross the tabulated values from the CDMS
catalogue \citep{cdms}. Accordingly, the CN/C$^{15}$N column density
ratio is proportional to the ratio of the integrated opacity scaled by
a species-dependent prefactor $f(ul,T_{\rm ex})$ which depends upon
the excitation temperature of each transition. In the following, we
demonstrate that a single excitation temperature can be adopted for
both sets of optically thin hyperfine lines of CN and C$^{15}$N. We
also estimated the uncertainty associated to this assumption.

On general terms, the excitation state of each molecule results from
the competition between radiative and collisional processes, which may
vary significantly within the disk as the density and kinetic
temperature depend on the radius $R$ and height $Z$ above the midplane
(see Fig.~\ref{fig:isolines}). However, it must be emphasized that our
measurement of the isotopic ratio does not require absolute
determinations of the column densities of each isotopologue. Instead,
it is important to guarantee that the single excitation temperature
assumption holds.

We examined in more details the assumption that all hf lines are
characterized by a single excitation temperature. This is supported by
the relatively small total opacity of each group of hf
lines. Moreover, the excitation temperature of the CN rotational
transition, as obtained from the \verb+HFS+ fits, are 17--27~K for
both fine structure groups (see Table~\ref{tab:hfs}). In our analysis,
spectra labelled 4 and 5 in Fig.~\ref{fig:hfsa}, for which the fits are
the poorest because the main hf transitions overlap, have the highest
opacities (total $\tau$ of 6.1). At any given location, the opacity of
each group are consistent to better than 15\%. The excitation
temperatures are also in very good agreement with the 25~K value of
the CN(2-1) lines at 100~AU \citep{teague2016}. This provides is a
very strong indication that the CN(2-1) and (3-2) lines are
thermalized, or very close to be so.  The derived excitation
temperatures are higher than in our single-dish study, confirming the
suggestion that the beam dilution was not properly taken into account
\citep{guilloteau2016}. The corresponding total CN column density is
$\approx10^{14}$~cm$^{-2}$, although we stress that a robust
determination of the CN column density should await for a
self-consistent modelling of the CN(2-1) and CN(3-2) lines observed
with ALMA.

The single excitation temperature assumption is further substantiated
by the FWHM of the CN lines which indicate that the emission
originates from regions with high density. At any location, the FWHM
measured through Gaussian fitting to each hf component, is due to the
combined effects of Keplerian motion and non-thermal (e.g. turbulence)
plus thermal broadenings, and therefore provides an upper limit on the
kinetic temperature through
$\sigma_v = {\rm FWHM}/2.35 \ge (2kT_k/\mu)^{1/2}$, with $\mu=26$~amu
for the CN molecule. {The maps of the FWHM of the \emph{group
    a} hf transitions (see Table~\ref{tab:spectro}) are shown in
  Fig.~\ref{fig:fwhm}. The associated upper limit on the kinetic
  temperature assuming purely thermal broadening are also indicated.}
The quoted FWHMs have not been deconvolved from the autocorrelator
transfer function. The Keplerian smearing associated to the $7^\circ$
disk inclination was also not subtracted, so that the upper limits on
$T_k$ are conservative ones. {The FWHMs of the optically thin
  hf lines at 340.008 and 340.020~GHz correspond to kinetic
  temperature lower than 20--25~K within the CN-ring (i.e. with
  convolved radii $\lesssim90$~AU, see Fig.~\ref{fig:azimuth}). These
  upper limits further decrease at larger radii, reaching
  $T_k \le 15$~K beyond 120~AU.}

\begin{figure}
  \centering
  \includegraphics[width=.9\hsize]{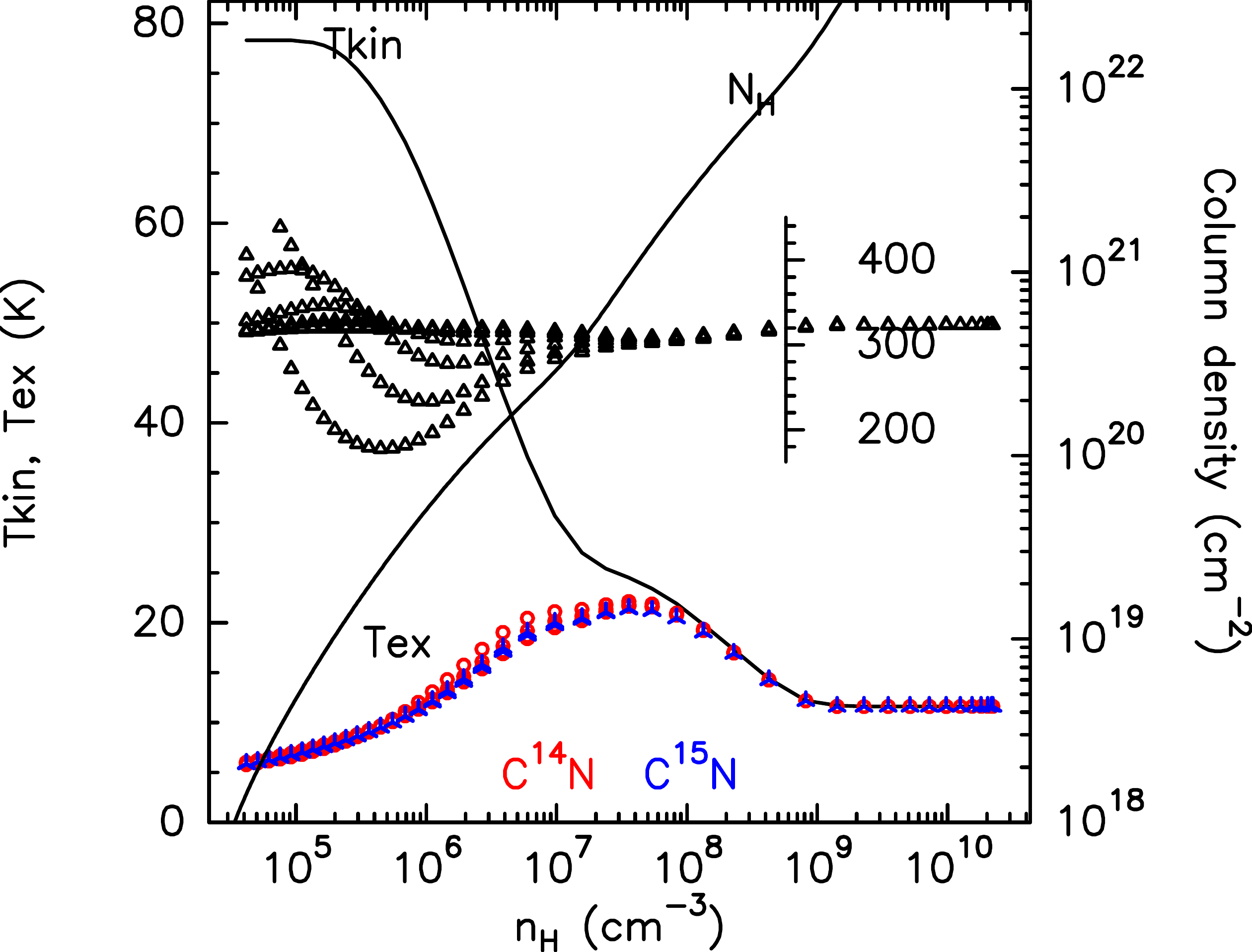}\bigskip\\
  \includegraphics[width=.9\hsize]{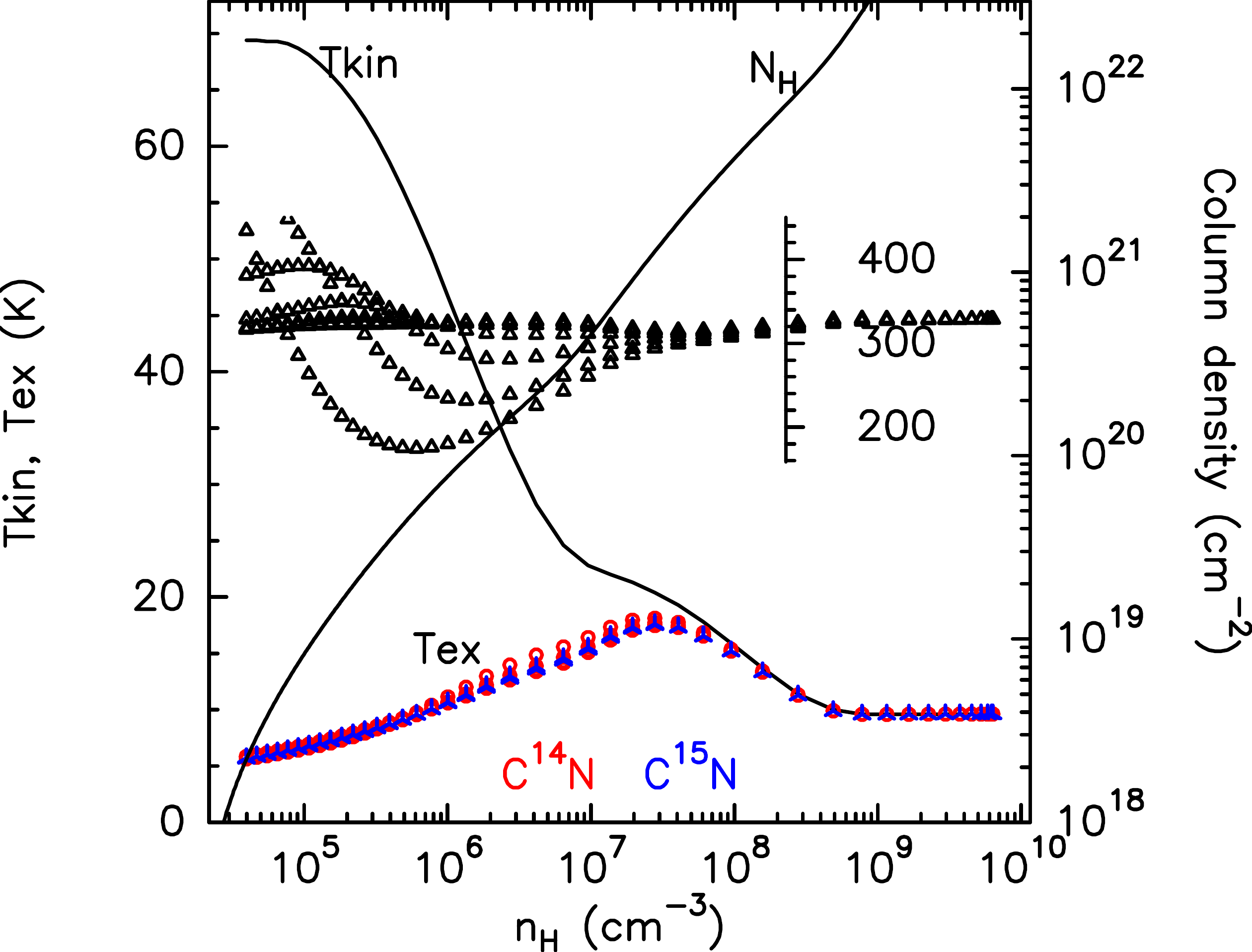}
  \caption{Physical structure and excitation temperatures of
    CN($\nu= 340.265$~GHz) and C$^{15}$N, {within the CN-ring,
      at radii of $R=60$ (top) and 85~AU (bottom)}. The total proton
    number density and kinetic temperature profiles are taken from
    \citep{qi2013}. The corresponding cumulative column density,
    increasing towards the midplane, is also shown (right scale). The
    excitation temperatures $T_{\rm ex}$ of the CN and C$^{15}$N
    hyperfine components (star triangles and open circles, resp.)
    considered in our $uv$-plane analysis (see Sections A.2 and A.3)
    were obtained in the escape probability formalism using the RADEX
    code \citep{vandertak2007}, and adopting a FWHM of 0.25 km/s. For
    each density and temperature, a range of CN total column densities
    were adopted, from 10$^{12}$ to 10$^{15}$ cm$^{-2}$, with
    C$^{15}$N being 300 times less abundant. Also shown is the
    CN/C$^{15}$N isotopic ratio (open triangles, inset scale) obtained
    by use of Eq.~\ref{eq:cdens} when adopting the CN and C$^{15}$N
    excitation temperatures from the non-LTE calculations.}
  \label{fig:tex}
\end{figure}

Based on a model of the physical structure of the TW~Hya disk
\citep{qi2013}, the {20--25~K} upper limit corresponds to a
lower limit on the density of {a few} 10$^7$~cm$^{-3}$ for
radii comprised in the range 60 to 120~AU. The non-equilibrium
statistical populations of the CN and C$^{15}$N hyperfine levels were
calculated under the escape probability formalism using the RADEX code
\citep{vandertak2007} and collision rates coefficients, for CN and
C$^{15}$N, at the hyperfine level \citep{hilyblant2013b}. At several
radii from 60 to 120~AU, level populations were computed as a function
of height above the disk midplane for physical conditions appropriate
to the TW~Hya disk. {Figure~\ref{fig:tex} shows the results at
  $R=60$ and 85~AU, which correspond to upper limits on $T_k$ of
  $\approx$25 and 20~K respectively}. At each density and kinetic
temperature, a range of CN column densities from 10$^{12}$ to
10$^{15}$~cm$^{-2}$ was explored while that of C$^{15}$N was a factor
300 or 441 lower. The value of the isotopic ratio was not found to
change the following results. At heights such that $T_k<20-25$~K, the
excitation temperature of the 340.020 CN line and the 329.837
C$^{15}$N line are equal to better than 3\%. At higher kinetic
temperatures, the density decreases and both transitions start to
deviate from thermalization, although very moderately. The present
excitation study includes locations well within the CN ring
(Fig.~\ref{fig:azimuth}), and is thus representative of the excitation
of the bulk of the CN emission. In addition to provide strong support
to the single excitation temperature assumption, our analysis suggests
that the CN emission does not originate in the upper, warm,
layer. Yet, 2D radiative transfer calculations are required to
demonstrate this claim.

\def\ha{0.33\hsize}
\def\hb{0.3\hsize}
\begin{figure*}
  \centering
  \includegraphics[width=\ha]{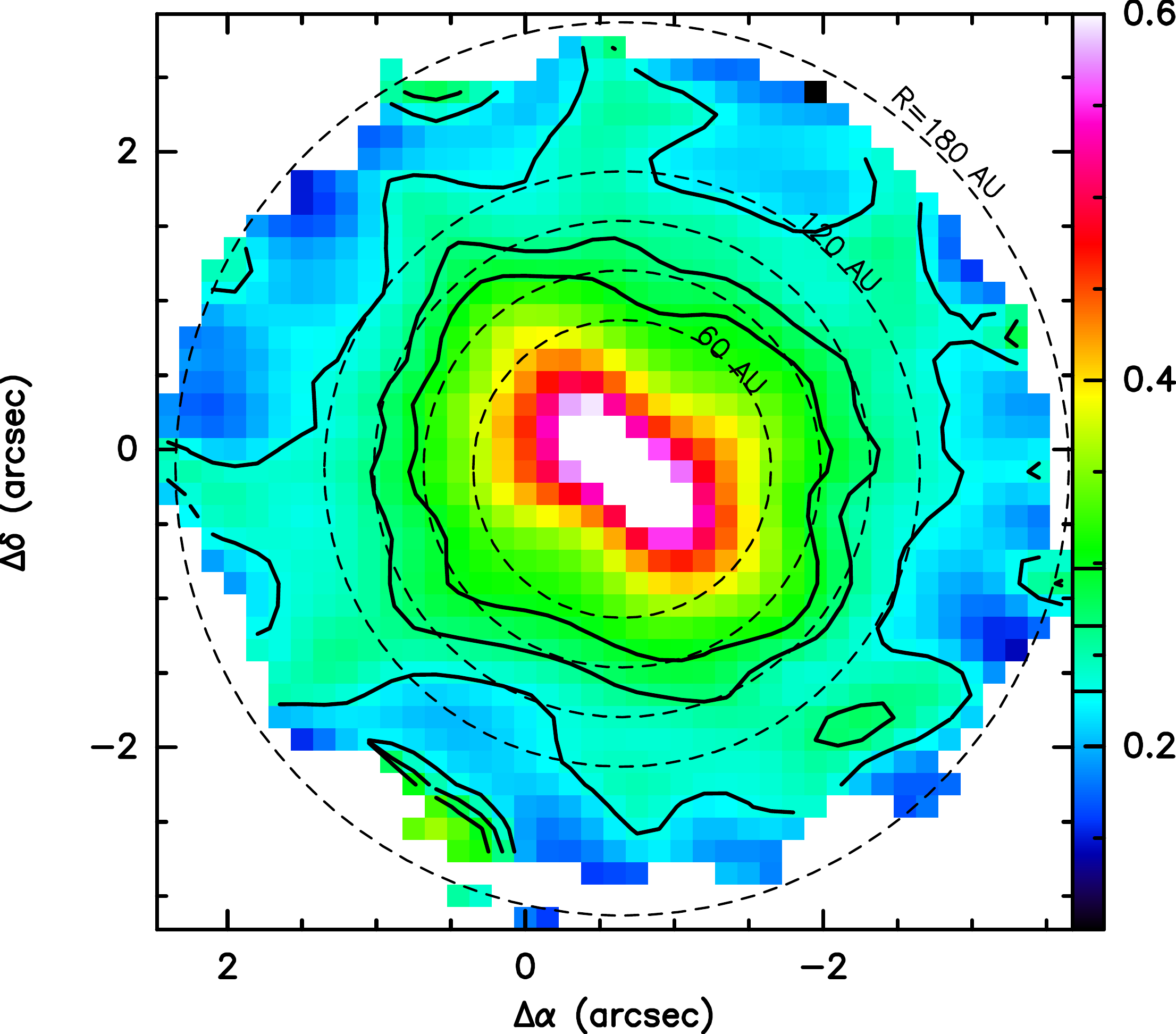}\hfill%
  \includegraphics[width=\ha]{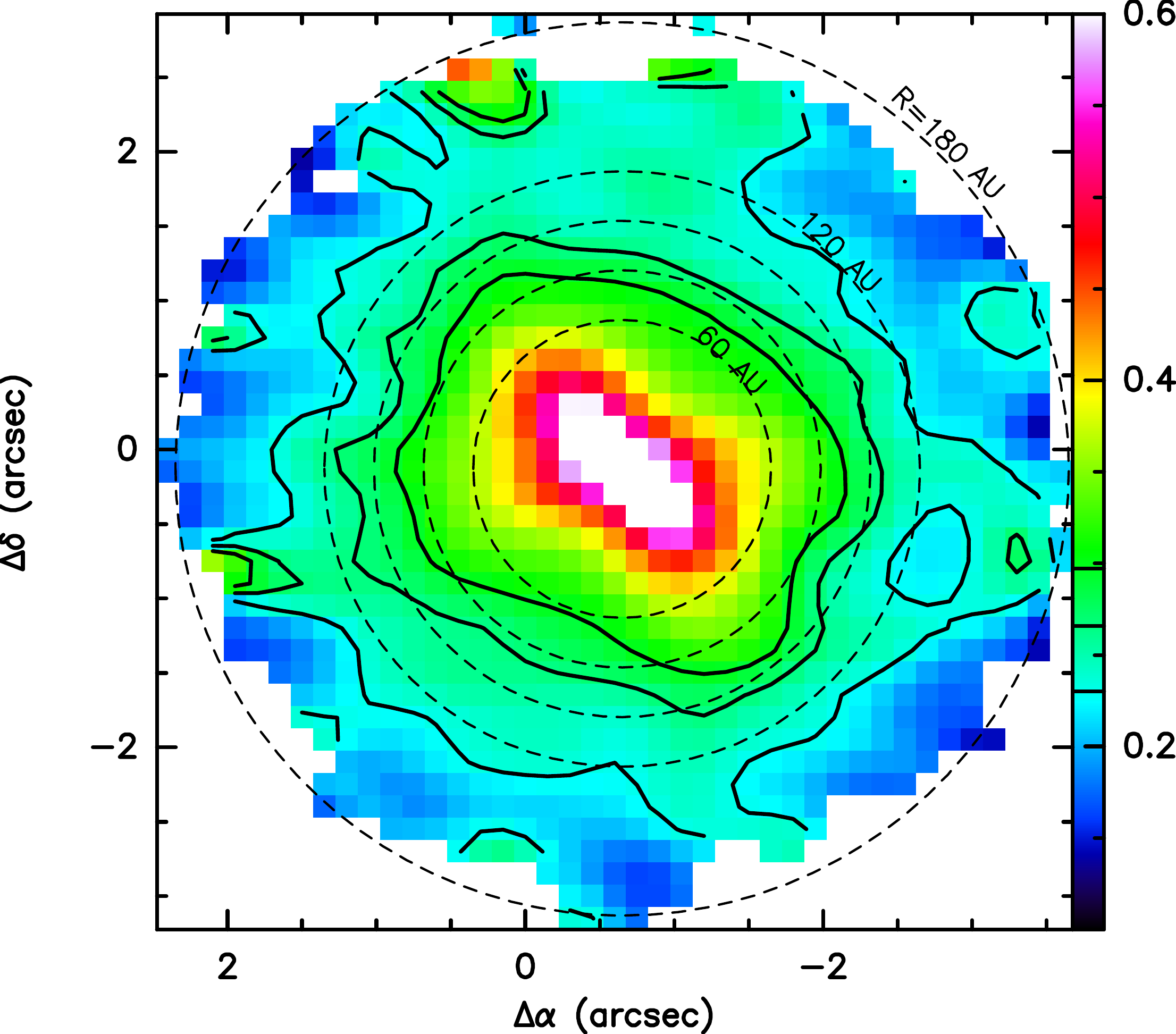}\hfill%
  \includegraphics[width=\ha]{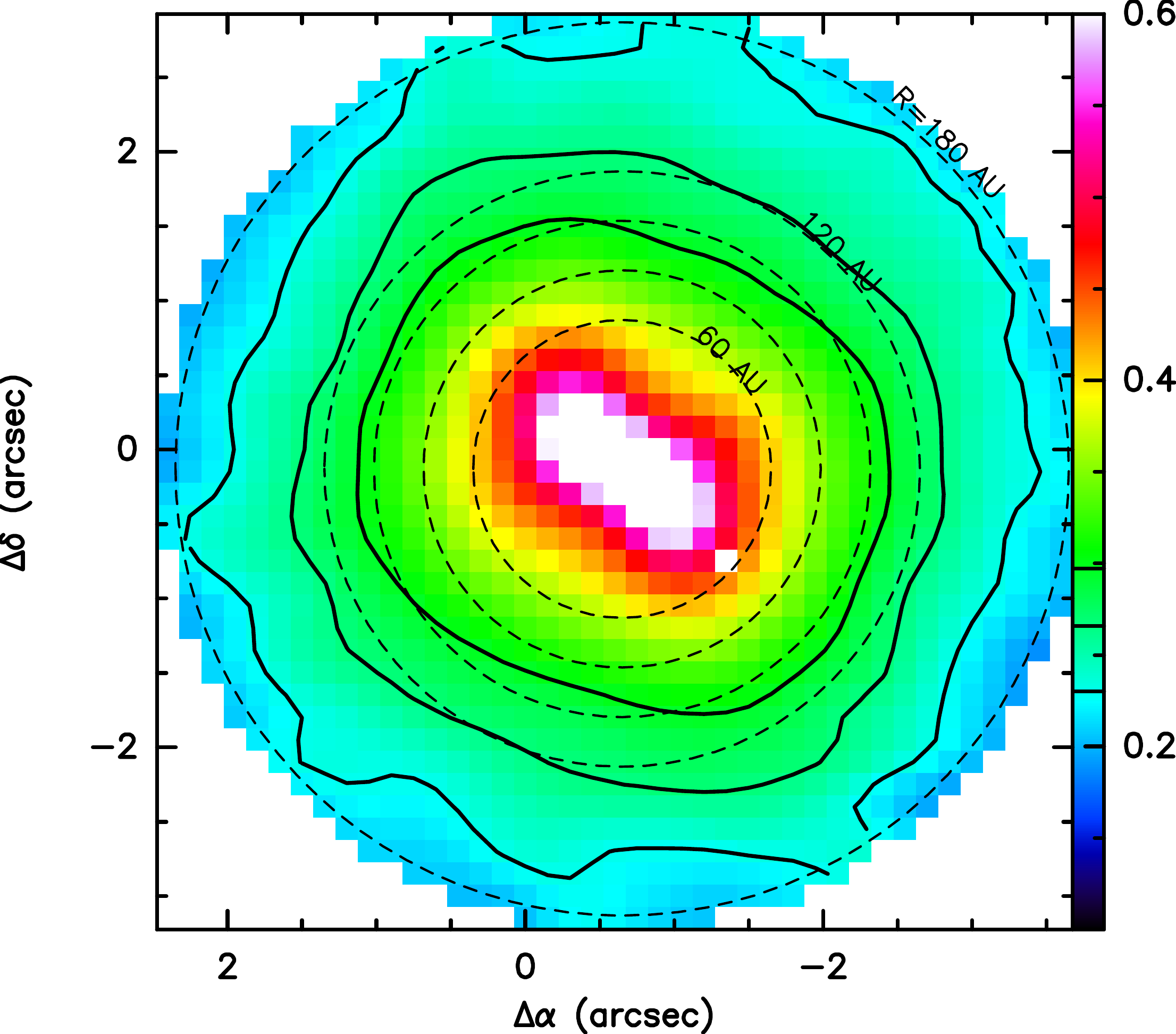}\bigskip\\
  \includegraphics[width=\ha]{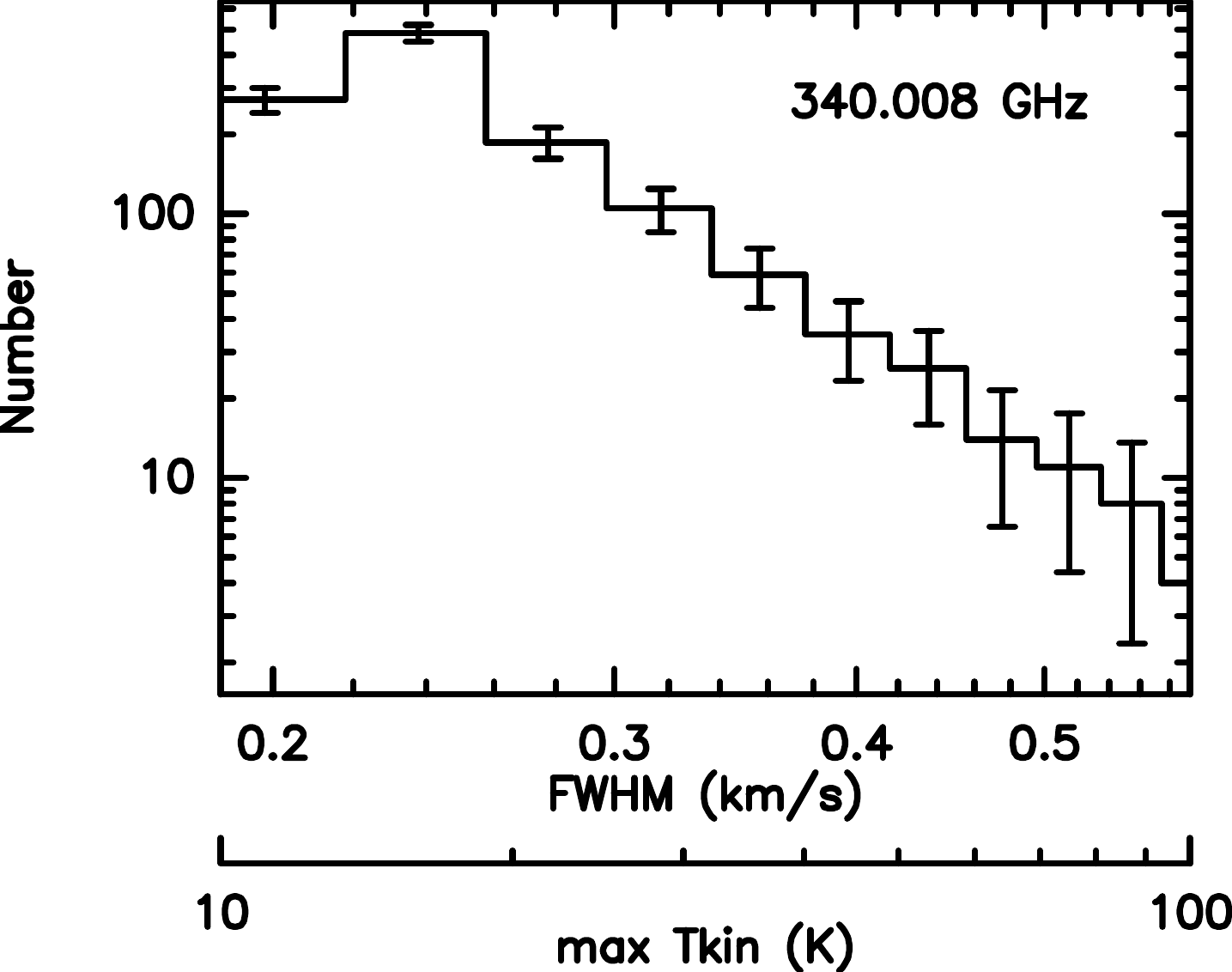}\hfill%
  \includegraphics[width=\ha]{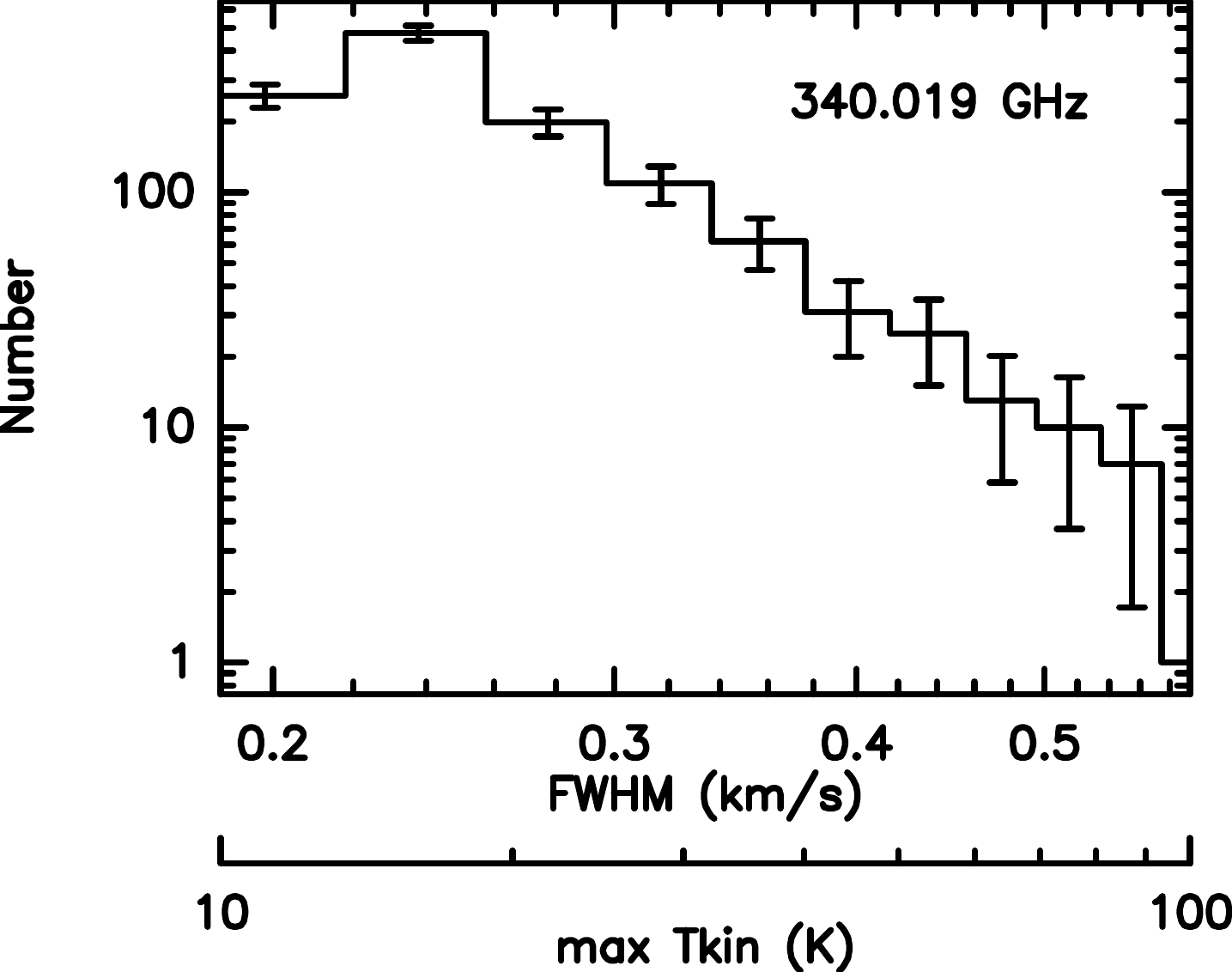}\hfill%
  \includegraphics[width=\ha]{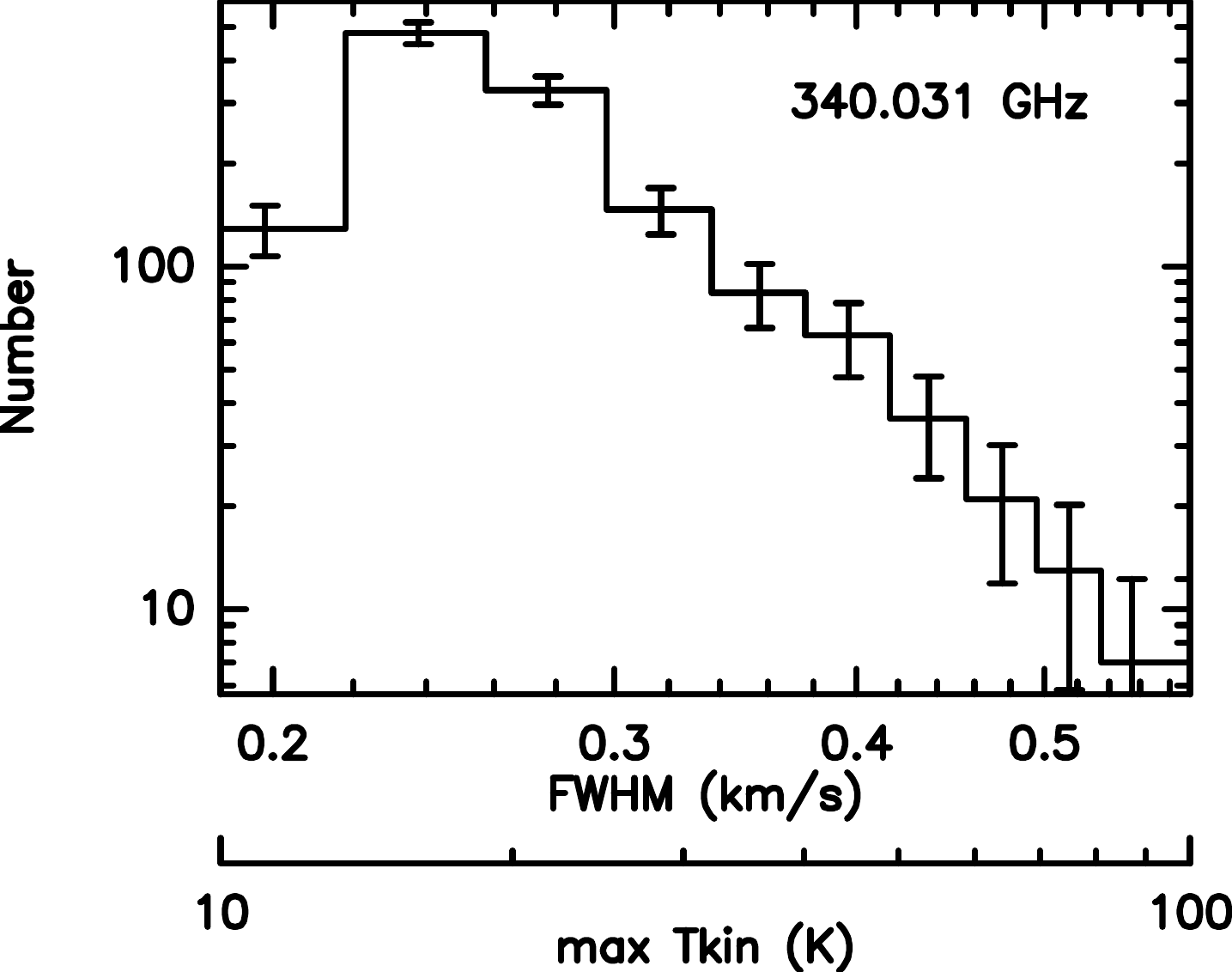}
  \caption{{Maps (top) and histograms (bottom) of the raw FWHM (in
      km/s) of three CN hf lines obtained from Gaussian fits to each
      spectrum. Only spectra with an area S/N larger than seven were
      kept. The contours (top panels) show the locus of constant
      kinetic temperature (15, 20, 25~K) adopting
      FWHM/2.35$=(2kT_k/\mu)^{1/2}$. The distance to the source is
      59.5~pc and the effective velocity resolution is 107 m/s.}}
  \label{fig:fwhm}
\end{figure*}

\begin{table}
  \begin{center}
    \caption{\label{tab:hfs}Results of the \texttt{HFS} fits (total
      opacity, and excitation temperature), for groups $a$ and $b$, at
      each location (see Fig.~\ref{fig:hfsa}).}
  \begin{tabular}{l c rr rr}
    \toprule
    \# & $R$ & $\tau_a$ & $T_{\rm ex, a}$ & $\tau_b$ & $T_{\rm ex, b}$\\
    & (AU) & & (K) & & (K)\\
    \midrule
    1 & 126.2 & 3.4  & 20 & 3.4 & 17 \\
    2 & 106.3 & 4.2  & 24 & 4.3 & 21 \\
    3 &  86.5 & 4.8  & 27 & 4.9 & 24 \\
    \it 4$^\dag$ &  \it62.0 & \it5.4  & \it32 & \it6.2 & \it28 \\
    \it5$^\dag$ & \it 74.4 & \it6.1  & \it29 & \it5.0 & \it27 \\
    6 &  94.2 & 5.4  & 25 & 4.3 & 23 \\
    7 & 101.8 & 5.1  & 24 & 4.1 & 22 \\
    8 & 121.8 & 4.5  & 20 & 3.2 & 19 \\
    \bottomrule
  \end{tabular}
\end{center} {\footnotesize $^\dag$ main hf components overlap.}
\end{table}

At each radius and height {with $T_k<20$~K}, the prefactors in
Eq.~\ref{eq:cdens} were finally computed, leading to a prefactor ratio
of
\begin{equation}
  f_{14}/f_{15}=26.5(8).
\end{equation}

The uncertainty in Eq.~\ref{eq:cnratio}, including all aforementioned
contributions, is finally dominated by the statistical uncertainties
on the C$^{15}$N visibilities:
\begin{equation}
\frac{\delta R}{R} =
\left[
\left(\frac{\delta (f_{14}/f_{15})}{f_{14}/f{_{15}}}\right)^2 
+
\left(\frac{\delta S_{14}}{S_{14}}\right)^2 
+
\left(\frac{\delta S_{15}}{S_{15}}\right)^2 
\right]^{1/2}
= 10\%
% values from the Table
% first estimate: [(0.8/26.5)^2+(0.02/0.94)^2+(0.01/0.10)^2]^{1/2} = 11\%
\end{equation}

% The prefactors were computed for CN(3 5/2 3/2 $\rightarrow$ 2 3/2 3/2)
% and C$^{15}$N. For the latter, it carefuly handles the averaging of the
% two hf transitions from the (3 7/2 7/2) upper level. A single
% excitation temperature was adopted for both isotopologues, with a
% value of 20~K based on the excitation analysis performed on our CN
% deconvolved spectra (Section A.3). Non-LTE calculations in the escape
% probability formalism using the publicly available RADEX code
% \citep{vandertak2007}, show that the single excitation temperature
% assumption is valid to within 5\%, over a wide parameter space with
% H$_2$ density from $10^3$ to $10^8$~cm$^{-3}$, kinetic temperature
% from 5 to 15~K, CN column densities from $10^{12}$ to
% $10^{16}$~cm$^{-2}$ and a CN/C$^{15}$N abundance ratio from 250 to
% 500. Since the prefactors depend on the species and on the excitation
% temperature, their ratio for CN and C$^{15}$N does not cancel the
% $T_{\rm ex}$ dependence. However, the spectroscopic properties of the
% CN and C$^{15}$N transitions are sufficiently close that the sensitivity
% to the excitation temperature between 15 and 30~K remains below 1.6\%.

\section{Fractionation in cores and disks}
\label{app:fractionation}

Here, we examine further the possibility that the consistency between
the isotopic ratio in CN in TW~Hya on the one hand, and in ammonia and
dyazenilium in prestellar cores on the other hand, represent the
elemental ratio rather than the result of fractionation
processes. Since the elemental ratio is the same in these objects, the
potential fractionation processes would have to be equally efficient
although under significantly different disk- and core-like physical
conditions. Indeed, disks have much larger density and are exposed to
higher UV fields compared to prestellar cores, which are exposed to a
higher flux of ionizing cosmic-rays than disks
\citep{henning2013,cleeves2014}. Hence, fractionation processes in
disks and cores, if efficient, are likely of a different nature,
respectively selective photodissociation or chemical mass
fractionation.

Chemical fractionation through gas-phase reactions is a
temperature-driven process, and for nitrogen, it is expected to become
efficient at temperatures significantly lower than 30~K
\citep{terzieva2000}. At large density ($>10^7$cm$^{-3}$), for
example, close to the disk midplane where considerable freeze-out of
gas-phase species takes place, species-dependent depletion may
strongly enhance fractionation \citep{charnley2002}, although the
resulting amount of ammonia in ices is outrageously large with respect
to observations of interstellar ices \citep{bottinelli2010} and even
comets \citep{mumma2011}. At the lower densities typical of prestellar
cores ($\sim 10^4$ cm$^{-3}$), the fractionation level may reach 25\%
on short timescales, and $\sim10\%$ at steady-state
\citep{terzieva2000, hilyblant2013b, roueff2015}. Regardless the
discrepancies between chemical model predictions in prestellar cores,
ammonia and dyazenilium are consistently found to be marginally
affected by fractionation. Selective photodissociation is also
negligible in such environments where the amount of
internally-generated UV photons is not sufficient to promote N$_2$
over N$^{15}$N \citep{heays2014}. Altogether, this strongly suggests
that NH$_3$ and N$_2$H$^+$ species are tracing the elemental
ratio. Nevertheless, measuring their isotopic ratios is not as direct
as for CN in TW~Hya, because radiative transfer effects can not be
neglected and spatial information, which is critical to disentangle
between line excitation and abundance gradients, is usually lacking
due to the intrinsic weakness of the $^{15}$N isotopologues
\citep{bizzocchi2013}. On the other hand, the total column density of
ammonia often relies on assumptions regarding its ortho-to-para ratio
\citep{lis2010}.

In contrast, models of UV-irradiated disks predict significant
enrichment of CN and HCN in $^{15}$N, at heights of 20 to 40~AU above
the midplane at $R=105$~AU, with CN/C$^{15}$N column density ratio
reaching 250 relative to an elemental ratio of 441
\citep{heays2014}. A simple scaling indicates that the \rrs{CN}=323 in
TW~Hya would require an elemental ratio of \rnow=575. However, such a
high elemental ratio {would disagree} with GCE models which
predict an enrichment in $^{15}$N with time, hence a present-day
elemental ratio lower than \rsun=441 (or lower than $\approx 500$ in
the outward Sun migration hypothesis). {This high value of
  $\rnow$ would also} require gas-phase NH$_3$ and N$_2$H$^+$ to be
significantly ($\sim 65-95$\%) fractionated in prestellar cores
envelopes, at odds with model predictions \citep{terzieva2000,
  hilyblant2013b, roueff2015}. Last, such a high elemental ratio would
also imply that CN and HCN are also strongly enriched in $^{15}$N in
diffuse molecular clouds, in which the ratios in these species are as
low as 250 (see Table~\ref{tab:ratios}), while the same model predicts
instead a depletion of $\approx 10$\% at most at visual exctinction
below 2~mag.

It appears more likely that the efficiency of N$_2$ selective
photodissociation has been overestimated in disk models. Indeed, we
note that these models predict CN and HCN column density ratios of 250
and 90 respectively, which are a factor of 1.3--1.4 smaller than the
ratios observed in TW~Hya and MWC~480. Yet, selective
photodissociation is primarily sensitive to UV propagation and hence
to the radial and vertical profiles of density and dust size
distribution and, probably to a lesser extent, to the kinetic
temperature and the ionization fraction of the gas. {To explain
  the observed CN and HCN ratios with these models and \rnow=441 would
  thus require to reduce} the efficiency of selective photodissocation
by $\approx 30$\%. {However, this would, in turn, imply that
  ammonia (\rr=321) is significantly enriched in $^{15}$N in
  prestellar cores (40\%)}, a possibility that models of fractionation
in prestellar cores consistently rule out. It appears that
contradicting consequences are obtained when assuming that the
isotopic ratio observed in disks and cores derive from a present-day
elemental ratio of 441, even when allowing for uncertainties on disk
and prestellar core fractionation models. Nevertheless, a definitive
proof that selective photodissociation has been overestimated in
published disk models requires thorough exploration of the wide
parameter-space of models selective photodissociation in disks is
timeley, with particular attention drawn on the impact of the dust
size distribution {and mass of the central protostar}.

\begin{table*}
  \begin{center}
    \caption{\label{tab:ratios} {Direct measurements of the} \nratio\
      isotopic ratio in the local ISM (at a distance $d$ to the Sun),
      derived from direct X$^{14}$N/X$^{15}$N abundance ratios using
      various molecular carriers and techniques. Uncertainties are at
      the 1$\sigma$ level. These data correspond to
      Fig.~\ref{fig:gce}.}
  \begin{tabular}{lllclrr}
    \hline
    Carrier    & Environment & Source    & Gal. coord.  & $d$ (pc)  & \nratio & Reference\\
    \hline
    HCN        & Diffuse    & B0415+379$\dag$ & 161.7, -8.8  & $<$1000 & 244(89)  & \citep{lucas1998}\\
    HCN        & Diffuse    & B0415+379$\dag$ & 161.7, -8.8  & $<$1000 & 282(37)  & \citep{lucas1998}\\
    CN         & Diffuse    & HD 73882        & 260.2,  0.64 & $<$1000 & 234(35)  & \citep{ritchey2015}\\
    CN         & Diffuse    & HD154368        & 350.0,  3.22 & $<$1000 & 452(107) & \citep{ritchey2015}\\
    CN         & Diffuse    & HD169454        &  17.5, -0.67 & $<$1000 & 283(22)  & \citep{ritchey2015}\\
    CN         & Diffuse    & HD210121        &  56.9,-44.5  & $<$1000 & $>312$      & \citep{ritchey2015}\\
%    \ce{NH3}  & Barnard 1 & 159.2, -20.1 & 235     & 334$\pm$20  & \citep{lis2010}\\
    NH$_3$     & Prestellar & Barnard 1       & 159.2, -20.1 & 235     & 334(17)  & \citep{lis2010}\\
    NH$_3$     & Prestellar & Barnard 1       & 159.2, -20.1 & 235     & 307(50)  & \citep{daniel2013}\\
    NH$_2$D    & Prestellar & Barnard 1       & 159.2, -20.1 & 235     & 255(80)  & \citep{daniel2013}\\
    N$_2$H$^+$ & Prestellar & I16923E         & 353.9,  15.8 & 125     & 365(135) & \citep{daniel2016}\\
    N$_2$H$^+$ & Prestellar & L1544           & 178.0,  -9.7 & 150     &1080(160) & \citep{bizzocchi2013}\\
    CN         & Disk       & TW Hya          & 278.7,  23.0 &  59.5   & 323(30)  & This work\\
    CN         & --- & Interpolation$\ddag$ & ---  &      0  & 290$\pm$40  & \citep{adande2012}\\
    \hline
  \end{tabular}
  \end{center}
  {\footnotesize {Notes:} $^\dag$ Two velocity components were
    detected towards this source.  $^\ddag$ Galactic gradient obtained
    from several source types.}
  \end{table*}

\end{appendix}  

\end{document}